\documentclass[DIV=12]{scrartcl}

\usepackage[arxiv]{optional}

\usepackage{amsmath}
\usepackage{amssymb}
\usepackage{amsthm}
\usepackage{thmtools,thm-restate}
\usepackage{forloop}
\usepackage{bm}
\usepackage{bbm}
\usepackage{algorithmic}
\usepackage{mdframed}

\usepackage{xspace}

\usepackage[l2tabu, orthodox]{nag}

\usepackage{hyperref}

\usepackage{nth}
\usepackage{color}

\usepackage{array}

\usepackage{paralist}
\usepackage{times}

\usepackage{subfig}
\usepackage{import}

\usepackage[capitalize, nameinlink]{cleveref}
\crefname{theorem}{Theorem}{Theorems}
\Crefname{lemma}{Lemma}{Lemmas}
\Crefname{claim}{Claim}{Claims}
\Crefname{observation}{Observation}{Observations}
\Crefname{invariant}{Invariant}{Invariants}

\usepackage{flafter}

\usepackage{tikz}
\usetikzlibrary{calc}
\usetikzlibrary{matrix}
\usetikzlibrary{shapes}
\usetikzlibrary{decorations.pathmorphing}
\usetikzlibrary{decorations.pathreplacing}

\newtheorem{theorem}{Theorem}[section]
\newtheorem{lemma}[theorem]{Lemma}

\newtheorem{Definition}[theorem]{Definition}

\newtheorem{corollary}[theorem]{Corollary}
\newtheorem{observation}[theorem]{Observation}

\newtheorem{claim}[theorem]{Claim}
\newtheorem*{claim*}{Claim}

\newtheorem*{fact*}{Fact}
\usepackage{xparse}
\theoremstyle{definition}

\colorlet{darkgreen}{green!50!black}

\newcommand{\defcal}[1]{\expandafter\newcommand\csname c#1\endcsname{{\mathcal{#1}}}}
\newcounter{ct}
\forLoop{1}{26}{ct}{
    \edef\letter{\Alph{ct}}
    \expandafter\defcal\letter
}

\newcommand{\E}{{\mathbb{E}}}

\newcommand{\hide}[1]{}
\newcommand{\sgraph}[1]{\ensuremath{\mathcal{SG}
\ifthenelse{\equal{#1}{}}{}{(#1)}
}}
\newcommand{\cgraph}[1]{\ensuremath{\mathcal{CG}
\ifthenelse{\equal{#1}{}}{}{(#1)}
}}
\newcommand{\cpath}[2]{\ensuremath{P_{#1}
\ifthenelse{\equal{#2}{}}{}{(#2)}
}}

\newcommand{\klogn}{k (\log \Delta+1)}
\newcommand{\GOPT}{{\ensuremath{\mathrm{GOPT}}}\xspace}
\newcommand{\OPT}{{\ensuremath{\mathrm{OPT}}}\xspace}
\newcommand{\epochalgo}{{\ensuremath{\mathrm{EpochAlgorithm}}}\xspace}

\newcommand{\dist}{{\ensuremath{\mathrm{dist}}}\xspace}

\newcommand{\avgcost}{{\ensuremath{\mathrm{avgcost}}}\xspace}
\newcommand{\Ball}{{\ensuremath{\mathrm{Ball}}}\xspace}

\newcommand{\makeHOAlg}{{\ensuremath{\mathrm{MakeRobust}}}\xspace}
\newcommand{\algrobustify}{{\ensuremath{\mathrm{Robustify}}}\xspace}

\newcommand{\cost}{{\ensuremath{\mathrm{cost}}}\xspace}

\newcommand{\aspratio}{\ensuremath{\Delta}}

\begin{document}

\title{\LARGE Consistent k-Clustering for General Metrics}
\author{
Hendrik Fichtenberger\footnotemark[1]
\and
Silvio Lattanzi\footnotemark[2]
\and
Ashkan Norouzi-Fard\footnotemark[3]
\and
Ola Svensson\footnotemark[4]}
\date{}

\maketitle

\renewcommand{\thefootnote}{\fnsymbol{footnote}}
\footnotetext[1]{University of Vienna, Faculty of Computer Science, Austria. Email: hendrik.fichtenberger@univie.ac.at}
\footnotetext[2]{Google Research, Zurich. Email: silviol@google.com}
\footnotetext[3]{Google Research, Zurich. Email: ashkannorouzi@google.com}
\footnotetext[4]{School of Computer and Communication Sciences, EPFL. Email: ola.svensson@epfl.ch. Supported by the Swiss National Science Foundation project 200021-184656 ``Randomness in Problem Instances and Randomized Algorithms.'' }
\renewcommand{\thefootnote}{\arabic{footnote}}

\begin{abstract}
Given a stream of points in a metric space, is it possible to maintain a constant approximate clustering by changing the cluster centers only a small number of times during the entire execution of the algorithm?

This question received attention in recent years in the machine learning literature and, before our work, the best known algorithm performs $\widetilde{O}(k^2)$ center swaps (the $\widetilde{O}(\cdot)$ notation hides polylogarithmic factors in the number of points $n$ and the aspect ratio $\Delta$ of the input instance). This is a quadratic increase compared to the offline case --- the whole stream is known in advance and one is interested in keeping a constant approximation at any point in time --- for which $\widetilde{O}(k)$ swaps are known to be sufficient and simple examples show that $\Omega(k \log(n \Delta))$ swaps are necessary. We close this gap by developing an algorithm that, perhaps surprisingly, matches the guarantees in the offline setting. Specifically,   we show how to maintain a constant-factor approximation for the $k$-median problem by performing an optimal (up to polylogarithimic factors) number $\widetilde{O}(k)$ of center swaps. To obtain our result we leverage new structural properties of $k$-median clustering  that may be of independent interest.
\end{abstract}


\newpage

\section{Introduction}
\label{sec:introduction}
Detecting the clustering structure of real-world data is
a basic primitive used in a wide range of data analysis
tasks such as community detection, spam detection, computational
biology and many others. Many different formulations of clustering problems
have been proposed and studied throughout the years. Among these,
the geometric versions of the problem has attracted a lot of
attention for their theoretical and practical importance.
In those problems we are given as input $n$ points in 
a metric space and a  distance oracle and we want to compute 
a clustering that minimizes an $\ell_p$-objective function, such 
as the $k$-means, the $k$-median or the $k$-center objective.

Due to their relevance, the problems have been extensively
studied and many algorithms~\cite{ahmadian2017better, arthur2007k, arya2004local, byrka2014improved, charikar2005improved, kanungo2004local, jain2003greedy, li20111} and
heuristics~\cite{lloyd1982least} have been proposed to solve
the classic offline version of the problem.
Furthermore they have been also extensively studied in
the classic streaming and online setting setting where points are inserted in the
instance sequentially. For those problems different techniques have been
used in the Euclidean $\mathbb{R}^d$ setting~\cite{chen2009coresets, feldman2007ptas, feldman2011unified, feldman2013turning, har2004coresets, har2007smaller} and in the
general metric space setting~\cite{charikar2003better, fotakis2011online, meyerson}.
In particular in the first case the algorithms are mostly based 
on coreset constructions. By contrast, in general metric spaces the algorithms 
are mostly based on adaptive sampling.

Inspired by real world applications where datasets continuously evolve in
time, we study the consistent $k$-median problem in general metric spaces where points are inserted on the fly. 
In this problem we are interested in efficiently maintaining a valid solution that is a good approximation of the
optimal solution and that is \emph{consistent}.
In particular given a stream of insertions, we are interested in designing an algorithm that maintains a constant-factor approximation at any time while minimizing the total number of changes in consecutive solutions.

Consistency of the solution is of importance from both a theoretical and a practical point of view and in recent
years it received a lot of attention in the machine learning literature~\cite{cohen2019fully, jaghargh2019consistent, 
lattanzi2017consistent}. In fact, in various applications the computed cluster centers are used in other downstream tasks.
In this case, changes in the solution might result in costly recomputations.

To formally capture the notion of consistency we use the same definition as in~\cite{lattanzi2017consistent}
where the consistency of a clustering is measured by the number of times the cluster centers are modified
during the entire execution of the algorithm. Intuitively, this definition makes sense because it captures the number
of times in which the clustering changes its underlying structure defined by its cluster centers. 

For the consistent $k$-clustering problem Lattanzi and Vassilvitskii~\cite{lattanzi2017consistent} presented a first
constant-factor approximation algorithm that executes at most $k^2 \cdot \text{polylog} (n, \Delta)$ changes, where 
$\Delta$ is the aspect ratio of input points. A simplified version of their algorithm intuitively works as follows: 
First they run the Meyerson's sketch \cite{charikar2003better,meyerson} to compress the stream down to $O(k \cdot \text{polylog}(n, \Delta))$ weighted point insertions and afterwards, for each point that is inserted 
in the weighted instance, they compute a solution using any constant-factor approximation algorithm. Moreover, they 
also show that, even if the stream of insertions is known in advance, any algorithm that maintains  $c$-approximate solutions requires at least
$\Omega (k \cdot (\log_c (\frac{n}{k})+\log_c(\frac{\Delta}{k}))$ changes. 
Finally, they also prove that if the stream of insertions is known in advance, one can design a constant-factor approximation
algorithm with $k \cdot\text{polylog} (n, \Delta)$ changes in the solution. Guo et~al.~\cite{guo2020consistent} study consistent $k$-clustering with outliers. Their algorithm is based on local search and requires at most $O(k^2 (\log n \Delta)^2)$ changes.

A natural open question left is to close the gap between the agnostic and the setting with knowledge of the future.
In this paper we solve this question by providing the first consistent $k$-median algorithm that maintains a constant-factor approximation at any point
in time by only changing the centers $k \cdot \text{polylog} (n, \Delta)$ times.

\paragraph{Our Results} We introduce a novel approach for the consistent k-median problem in metric spaces. 
We present a constant-factor approximation algorithm with a total of $k \cdot\text{polylog} (n, \Delta))$ changes in the consecutive 
solutions. Our approach is tight (up to a polylogarithmic factor) due to the above explained lower-bound. 
Moreover, surprisingly, our result shows that knowing the future is not a key information for this problem by achieving
the same number of changes as in the offline setting with full knowledge of the future (up to a polylogarithmic factor). 
To obtain our result we leverage new structural properties of $k$-median clustering and we introduce a new concept of 
robust centers that may be of independent interest. Maintaining a robust solution enables us to change the solution 
a little even for adversarial point insertions.
More precisely, in our algorithm, we first use the Meyerson's sketch \cite{meyerson} and get a stream of  
$k \cdot\text{polylog} (n, \Delta)$ weighted points. Afterwards we show that for each insertion we change 
the center of the maintained robust solution at most $\text{polylog} (n, \Delta)$ times on average, instead of 
$k \cdot\text{polylog} (n, \Delta)$ of the previous work. In order to achieve this, we establish a strong 
relationship between the input points and the metric space. We  change the position of the centers of the 
solution so that they have a better local coverage despite the fact that this might increase the cost of the clustering. 
This enables us to be more consistent in the future, whilst it increases the solution cost by a small factor.

\paragraph{Extensions to Other Problems.} For simplicity,  we state our result
for the classic $k$-median problem, although it is not hard to extend our results for
general $p$-norms for constant $p\geq 1$. In fact, it has already been noted that the
Meyerson sketch can be extended to this setting~\cite{lattanzi2017consistent} and our
reductions only use basic geometric properties as triangle inequalities so all our proofs can be adapted to work in general $p$-norms. Furthermore, for the same reason,
our result can also be extended to work in $\lambda$-metrics\footnote{$\lambda$-metrics are 
metrics where the triangle inequality hold only with an additional multiplicative $\lambda$ factor.} for constant $\lambda$.

\paragraph{Further related works} Interestingly, we note here that our notion of consistency is also closely related 
to the notion of recourse in online algorithms. 
In this setting, one seeks better online algorithms to compute optimal or approximate solutions for combinatorial 
problems by allowing the  algorithm to make a limited number of changes to the online solution. The first problem 
studied in this setting was the classic online Steiner tree problem introduced by~\cite{imase1991dynamic} 
for which it is possible to design better algorithms by allowing a small recourse as shown in several papers
\cite{gu2016power, gupta2014online, lkacki2015power, megow2012power}. 
After this result several other classic optimization problems have been studied in this setting as online scheduling
\cite{andrews1999improved, epstein2014robust, phillips1993online, sanders2009online, skutella2010robust}, online flow 
\cite{gupta2014maintaining, westbrook2000load}, online matching
\cite{bernstein2019online} and online set cover \cite{gupta2017online}.
\section{Overview of Our Approach}

The starting point of our approach are two basic observations that allow us to reduce our problem to an equivalent
simpler formulation that we will then solve efficiently. This reduction is not new, in fact it has been first
observed in~\cite{lattanzi2017consistent} (although in this paper we need to strengthen it slightly for our result).

The first observation is that we can divide the stream in $\log(n\Delta)$ phases so that in each phase the cost of 
an optimal solution does not change too much. In particular, after each insertion we compute an $\alpha$-approximate 
solution (for $\alpha  \leq 3$, using for example~\cite{byrka2014improved}). Now, when the current approximation 
cost increases by a factor $6$ in comparison to the cost at the beginning of a phase, we know that the value of 
the optimum solution has increased by at least a factor $2$ and at most a factor $18$ so we restart a new phase.
The second is that  within a phase with high probability we can reduce the problem with $n$ point insertions
to a problem where only $k \cdot \text{polylog}(n,\Delta)$ weighted points are inserted to the instance. This can be achieved
using a standard sketching tool, the Meyerson's sketch~\cite{charikar2003better,meyerson}.
This technique results 
in losing a constant-factor in the approximation ratio but it provides a way to transform any constant-factor approximation 
algorithm that makes on average $C$ changes to the cluster centers for each weighted point insertion to a constant
approximation an algorithm that makes $Ck \cdot \text{polylog}(n,\Delta)$ changes in total. Note that this automatically
results in a constant approximation algorithms that makes $k^2 \cdot \text{polylog}(n,\Delta)$ and it is the key idea behind
the algorithm presented in~\cite{lattanzi2017consistent}.

Now we can turn our attention to the main technical contribution of the paper. Assuming that the cost of the 
optimal solution is stable we want to design  an algorithm that maintains a constant approximation and makes on average 
$\text{polylog}(n, \Delta)$ changes to the cluster centers to handle insertions of weighted points.

Let $\mathcal{V}$ be the set of centers in the optimal clustering at the end of the phase and let $\mathcal{U}$
be the set of cluster centers computed by the algorithm. Now suppose that at any point in time during the execution 
of a phase we have that every center in $\mathcal{U}$ can be paired with a center in $\mathcal{V}$ so that the paired 
centers are close to each other but they are far away from all the other centers. Then in this setting we would expect 
future point insertions in the phase to modify the clustering structure only minimally. Intuitively,
this is true because the cost of the optimal solution does not increase too much and so the clustering induced by
the centers in $\mathcal{U}$ is fairly stable and centers exchange only points in their peripheries. In fact, in this
setting we can essentially reduce the problem to solving a set of disjoint $1$-median problems.

Unfortunately this approach does not generalize to more complex instances where the clustering structure is not
as neat. In fact, at any point in time we will have that only a (possibly empty) subset of points in $\mathcal{U}$
can be paired with a center in $\mathcal{V}$ so that the paired centers are close to each other but they are far away
from all other centers. We refer to those pairs as well-separated pairs (for a formal definition please refer to 
\cref{def:well-separated}). To tackle this more challenging setting we prove a few key structural lemmas.
From a high level perspective, our first key observation is that if the current set of centers is composed by $k-\ell$ 
well-separated pairs then it is possible to obtain another set of centers that has cost a constant time larger than 
the initial set of centers but that uses $\Omega(\ell)$ less centers. Informally, we prove such a statement by carefully 
constructing a fractional solution that opens all centers forming well-separated pairs and centers 
that are close to many optimal centers completely but opens only fractionally centers that are not close to multiple optimum centers (for more details refer to \cref{sec:number_wellseparated}). Intuitively, this is true because we can show
that centers that are not part of well-separated pairs have other centers in their proximity and so we can charge
points in their cluster partially to other centers.

Thanks to the previous observation we can design an algorithm as follows: given a current solution reduce the
number of centers as much as possible without increasing the cost of the induced clustering more than a constant factor.
Suppose that in this way, we can reduce the number of centers by $\ell'\in \Omega(\ell)$. After this step, we can easily insert $\ell'$ 
new points and open them as new centers. In this way we can handle $\ell'$ insertions without increasing the cost of
the solution. Unfortunately we cannot use this algorithm on its own to solve the problem because $\ell'$ may be $0$
and the cost of the solution may increase too much in subsequent iterations. To tackle this issues we prove our second
structural lemma where we show that after adding $\ell' + 1$ points to the current instance we can find a set of
$O(\ell + 1)$ centers swaps such that the solution obtained after those swaps is a $\beta$-approximation for the optimal
clustering for a fixed constant $\beta$. The main idea to prove such a lemma is to show that one can construct a $\beta$-approximation by keeping all the centers that are in well-separated pairs and for which no additional point has been
added to their induced cluster, and by swapping all other centers. One key idea in this context is to identify a set of
\emph{robust} centers for the clusters. The main idea behind this notion is that we want our centers to
be good centers for our clustering at different scales\footnote{Interestingly, we note that this notion is somehow
related to the notion of center used for prefix clustering in~\cite{mettu2003online}.} so that small changes in the 
periphery of our clusters do not affect their quality (for a formal definition and more details please refer to 
Section~\ref{sec:robust}). Interestingly, we can show in Section~\ref{sec:robust} that for any cluster it is possible
to find a center that is a ``good'' center for the cluster at any scale. Furthermore in Section~\ref{sec:robustifications}
we then show that those centers are robust meaning that they do not need to be updated often unless their clusters
change significantly.

Combining these two observations we can define our update algorithm, \epochalgo. It is called repeatedly in each phase
until the phase is finished. \epochalgo starts with a set of centers $\mathcal{U}$ such that: i) $\mathcal{U}$ has a constant
approximation ratio for the current set of points $P$, ii) all centers in $\mathcal{U}$ are \emph{robust}.
Then \epochalgo removes from $\mathcal{U}$ as many center as it can without significantly increasing the cost of the clustering. This set can be executed using an LP based algorithm (for more details
please refer to \cref{sec:removing-centers}). Suppose that the algorithm removes $\ell'$ centers. For the next $\ell'$ 
insertions, the algorithm inserts the inserted points as centers in the instance. Then an additional point is added, \epochalgo
finds $O(\ell)$ good centers swaps and performs them using another LP-based algorithm described in \cref{sec:swapping-centers}.
In this way we obtain a new good solution and the last step of \epochalgo is to make the new solution robust by robustifying the centers.

\paragraph{Roadmap.} We start by introducing some basic notation in \cref{sec:preliminaries}. Then we introduce the notion
of robust center and few basic properties of robust centers in \cref{sec:robust}. After formally defining the notion of robust
center we give a formal description of our \epochalgo in \cref{sec:epoch}. Then in \cref{sec:robustifications} and
\cref{sec:cost-of-epoch} we bound the number of changes in cluster centers and the approximation factor for
\epochalgo, respectively. Then in \cref{sec:everything_alg} we present our complete algorithm (with the preprocessing steps) and in \cref{sec:lp-algo} our algorithms for
reducing the number of centers and finding a good set of swaps. \opt{arxiv}{Finally, for completeness, in \cref{sec:meyerson} we show
how the Meyerson sketch can be used in our problem to reduce the size of the input instance.}\opt{soda}{In the full version, we show how the Meyerson sketch can be used in our problem to reduce the size of the input instance.}

\section{Preliminaries and Formal Statement of the Main Result}
\label{sec:preliminaries}
\newcommand{\Alg}{\ensuremath{\textsc{Alg}}}

In the consistent $k$-median problem, we are given a stream $\sigma = \langle \sigma_1, \sigma_2, \ldots, \sigma_n\rangle$ consisting of $n$ insertions of points and we are interested in maintaining a good and consistent solution to the $k$-median problem at any point in time. More formally, we want to maintain a solution that is a constant-factor approximation  while the number of changes in the solution is minimized throughout the execution of the algorithm.\footnote{We emphasize that the algorithm has no knowledge of the points $\sigma_{i+1}, \sigma_{i+2},  \ldots, \sigma_n$ at time $i$.} 

Given any two points $x_1$ and $x_2$, we assume to have access to a metric distance oracle that returns the distance between $x_1$ and $x_2$. 
We denote this distance by $\dist(x_1,x_2)$.
By scaling,  we assume that the smallest non-zero distance is $1$ and we let $\Delta$ denote the largest distance. We thus have that $\Delta$ equals the ratio between the largest distance to the smallest non-zero distance, which is often referred to as the aspect ratio. 
For a set $S$ of points and a point $x$, we also let $\dist(x, S) = \min_{y\in S} \dist(x,y)$ be the distance from $x$ to the closest point in $S$. By convention, $\dist(x, \emptyset) =\Delta$, i.e., we let the distance from a point to the empty set equal the largest possible distance. We remark that our algorithm does not needed to know the value of $n, \Delta$ in advance although for the sake of simplicity, we assume that we know these values (we explain how one can remove such assumption in \cref{sec:everything_alg}).
 
Given a set $P$ of points and a subset $\mathcal{U} \subseteq P$ of at most $k$ points, we refer to the points in $\mathcal{U}$ as centers.
A clustering is an assignment of points to centers.
For a set of centers $\mathcal{U}$ and a set $P$ of points, we let the cost of the clustering induced by $\mathcal{U}$ be
\begin{align*}\cost(\mathcal{U}, P) = \sum_{x\in P} \dist(x, \mathcal{U})\,. \end{align*}
In the case that the input points $P$ are weighted, by $w(x)$ we denote the weight of the point $x$ for $x \in P$ and we define the cost of the clustering induced by $\mathcal{U}$ to be 
\begin{align*}\cost(\mathcal{U}, P) = \sum_{x\in P} w(x) \cdot \dist(x, \mathcal{U})\,. \end{align*}
We also extend the notion of weights to sets, i.e., $w(P) = \sum_{x \in P} w(x)$. The $k$-median problem on input $P$ asks for a set $\mathcal{U}$ of $k$ centers that minimizes $\cost(\mathcal{U}, P)$ among all sets of $k$ centers.
We say that an algorithm $\Alg$  maintains an $\alpha$-approximate clustering, if at any point in time the clustering induced by $\mathcal{U}$ computed after an insertion has cost at most $\alpha$-times the optimal solution to the $k$-median problem.
Formally,  let $P_i$ be the set of points in our instance after the first $i$ insertions of the stream, i.e., $P_i = \{ \sigma_1, \ldots, \sigma_i\}$. Then we say that $\Alg$  maintains an $\alpha$-approximate solution if 
\begin{align*}
    \cost(\mathcal{U}_i, P_i) \leq \alpha\cdot  \OPT(P_i) %
    \qquad \mbox{for all $i=1,2,\ldots, n$,}
\end{align*}
where $\mathcal{U}_i$ denotes the centers computed  by $\Alg$ after the $i$-th insertion and  $\OPT(Q)$ denotes  the value of an optimal clustering of points in $Q$. 

The main focus of this paper is in developing an algorithm that maintains a constant approximation at any point in time by keeping the set of centers as stable as possible. This notion of stability is formally captured by the consistency of the center set, which is the summation of the number of changes in the consecutive sets of centers produced by the clustering algorithm. Formally, 
$$ \sum_{ 1 \leq i \leq n-1} |\mathcal{U}_{i+1} \setminus \mathcal{U}_{i}|.$$
In this work we present a constant approximate algorithm with $k \cdot \text{polylog} (n,\Delta)$ number of changes in the solution, matching the lower bound on the number of changes up to polylogarithmic factors.
\begin{restatable}{theorem}{mainresult} \label{thm:mainresult}
There exists an algorithm that given a stream of points $\sigma = \langle \sigma_1, \sigma_2, \ldots, \sigma_n\rangle$, outputs a set of at most $k$ centers $\mathcal{U}_i$ after the $i$-th insertion for all $1 \leq i \leq n$ so that with probability at least $1-\frac{1}{n^8}$
\begin{enumerate}
    \item $\mathcal{U}_i$ induces a constant approximation clustering, i.e., for any $1 \leq i \leq n$
    \begin{align*}\cost(\mathcal{U}_i, \{\sigma_1, \sigma_2, \ldots \sigma_i\}) \leq O(1)\cdot\OPT(P_i)\,.  \end{align*}
    \item The total number of changes in consecutive center sets is at most  $k \cdot \text{polylog} (n\Delta)$.
    \begin{align*}\sum_{1 \leq i < n} |\mathcal{U}_{i+1} \setminus \mathcal{U}_i| \leq k \cdot \text{polylog} (n, \Delta))\,. \end{align*}
\end{enumerate}
\end{restatable}

\section{Robust Centers}\label{sec:robust}

\begin{figure}[h]
    \centering
    \subfloat[]{%
        \label{fig:rd-motivation}%
        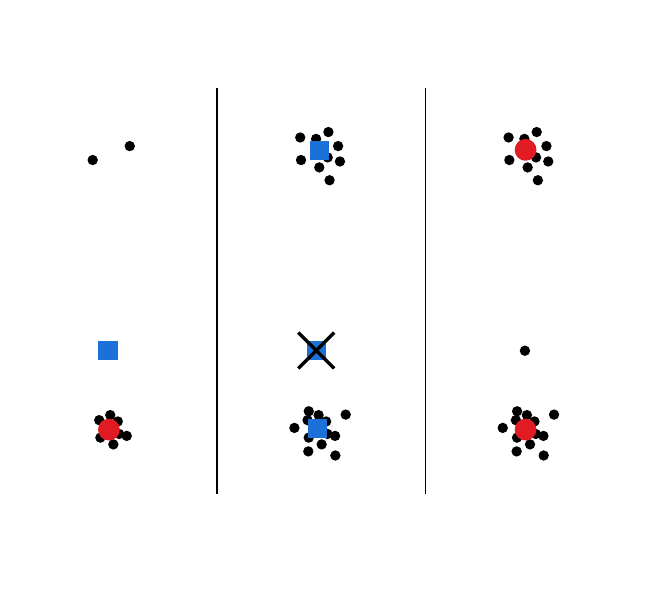%
    }%
    \qquad
    \subfloat[]{%
        \label{fig:rd-balls}%
        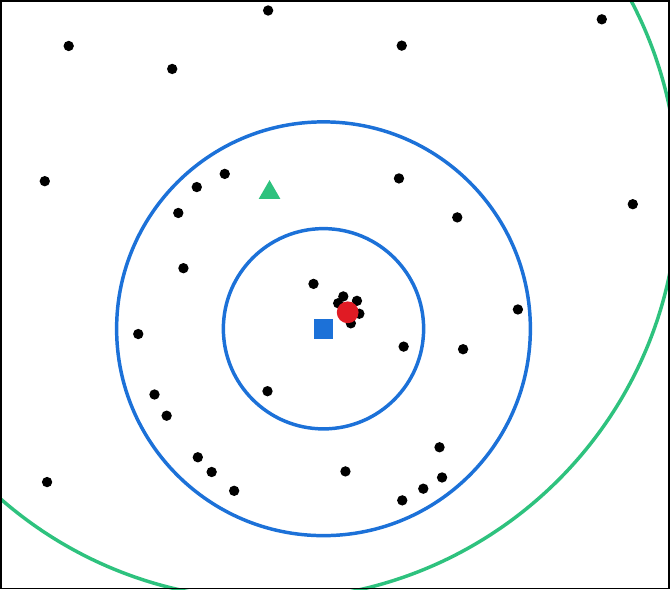%
    }
    \caption{%
        \protect\subref{fig:rd-motivation} Consider the two initial choices for a single center in the left clustering (red circle, blue square). Both are good choices and, depending on the metric, the blue center might have the smaller cost, but the red center is robust. However, if points are inserted and points accumulate in the upper \emph{and} the lower part, it may be necessary to split the cluster. In this case, the red center is better off in terms of consistency: while the blue solution needs to change two centers (the initial blue center and a new center, depicted in the middle), the red solution only needs to add a center at the top (depicted on the right); the original red center is already a good choice. %
        \protect\subref{fig:rd-balls} Consider a (sub-)sequence of a robust tuple: $p_i$ (green triangle), $p_{i-1}$ and $p_{i-2}$ (blue square), and $p_{i-3}$ (red circle). Since $\Ball(p_{i-1}, 10^{i-1})$ (second inner blue circle) has large average cost, we have $p_{i-1} = p_{i-2}$. The $\Ball(p_{i-2}, 10^{i-2})$ has small average cost, so $p_{i-2}$ is chosen optimally inside it.}
    \label{fig:robust-definition}
\end{figure}

One of the key concept in our analysis is the notion of \emph{robust} centers. Robust centers are not necessarily optimal centers, but they are robust to changes in the cluster structure. This  notion is central in our algorithm, as we aim to find a stable set of $k$ centers. To get an intuition behind our definition, consider a cluster in an approximately optimal solution that at a specific point in time has a rather blurry structure, e.g., many points in different location are good candidate centers. After few additional insertions the cluster structure may develop and concentrate, for example the cluster may grow, shrink, merge or split, but at its current state, it is not clear what will happen. In this setting an optimal solution with knowledge of future insertions would pick a center that is good for the current point set \emph{and} good after inserting future points (see \cref{fig:rd-motivation}\opt{soda}{ on page \pageref{fig:rd-motivation}}). Unfortunately, our algorithm cannot predict how an optimal consistent clustering would look without knowing the future. Nevertheless, we can select our centers so that they are stable to small changes in the clusters. To this end, we say that a center is robust when it is locally optimal on different scales, i.e., we optimize centers so that they minimize the assignment cost of close points better than the assignment cost of points far away. Intuitively, this makes sense because points that are far away from the center are potentially prone to moving between clusters. Again, an algorithm cannot tell a priori what \emph{far away} means because the scale of concentration of a cluster is unknown. Therefore, robust centers are optimized for (exponentially) growing distances.

Before formalizing this intuition we introduce few basic definitions.

\begin{Definition}
    For a center $u$ and a set of points $P$, we define the average cost of assigning the points in $P$ to $u$ by
    \begin{align*}
        \avgcost(u, P) = \frac{\cost(u, P)}{w(P)}\,.        
    \end{align*}
\end{Definition}

\begin{Definition}
    For a set of points $P$,  a point  $u\in P$,  and $r\geq 0$, we define $\Ball_P(u, r) = \{p\in P: \dist(u,p) \leq r\}$ to be the points of $P$ contained in the ball of radius $r$ centered at $u$. 
\end{Definition}
When the set $P$ of points is clear from the context, we simplify notation and write $\Ball(u, r)$ instead of $\Ball_P(u,r)$.

Let $\mathcal{U}$ be a set of centers for a point set $P$, for $u \in \mathcal{U}$ consider a subset of points that is assigned to this center, say, $P' = \Ball_P(u, \dist(u, \mathcal{U} \setminus \{ u \}) / 2)$. Now, imagine that after few insertion the set of centers changes and so the ball's radius decreases. Now we may need to move the center within the ball to a new optimal position, furthermore we can imagine that this could happen multiple times during the execution of the algorithm. Intuitively (e.g., for Euclidean distances), a center is attracted by accumulations of points in its current ball. To obtain a robust center we define our centers recursively so that they are good centers at different scale and so that they are not sensitive to accumulation of points in the boundary of the current ball. To do this we first discretize the shrinking process of the ball and consider exponential steps when decreasing its radius. Then we do not move the center when the average cost indicates that many of the points within the ball lie close to its boundary. This case indicates that, intuitively, there is no concentration of points close to the center that attracts it. See \cref{fig:rd-balls}\opt{soda}{ on page \pageref{fig:rd-balls}} for an example. More formally,

\begin{Definition}
    For an integer $t\geq 0$, we say that a tuple $(p_0, p_1, \ldots, p_t)$ of $t+1$ points is \emph{$t$-robust} with respect to a point set $P$ if, for every $i\in \{1,2,\ldots, t\}$, 
    \begin{align*}
        p_{i-1} & = \begin{cases} p_{i}
            \opt{arxiv}{&}\opt{soda}{\\\quad}
            \mbox{if $\avgcost(p_i, \Ball_P(p_{i}, 10^{i})) \geq 10^i/5$,} \\
        \arg\min_{p\in \Ball_P(p_i, 10^i)} \cost(p, \Ball_P(p_i, 10^{i}))
            \opt{arxiv}{&}\opt{soda}{\\\quad}
            \mbox{otherwise.}
        \end{cases}
    \end{align*}
    We also say that a point/center $u \in P$ is $t$-robust if there is a tuple $(u, p_1, \ldots, p_t)$ that is $t$-robust.
    \label{def:hierarchically_optimal}
\end{Definition}
Interestingly toward the end of this section we will prove that $p_0$ is an approximately good center for any ball $\Ball_P(p_i, 10^{i})$.

For simplicity, we also write that $(p_0, p_1, \ldots, p_t)$ is $t$-robust without explicitly stating that it is with respect to $P$ when there is no ambiguity. In particular, the point set will be fixed to be $P$ throughout this section.

Before proving two basic properties of $t$-robust tuples,  we describe a simple algorithm $\makeHOAlg(t,p)$ that, on input an integer $t\geq 0$ and a point $p\in P$, outputs a $t$-robust tuple $(p_0, p_1, \ldots, p_t)$ with $p_t = p$. 
 The steps of $\makeHOAlg(t,p)$ are as follows:
\begin{itemize}
    \item Let $p_t = p$.
    \item For $i$ from $t$ to $1$, select $p_{i-1}$ such that 
    \begin{align*}
        p_{i-1} & = \begin{cases} p_{i}
            \opt{arxiv}{&}\opt{soda}{\\\quad}
            \mbox{if $\avgcost(p_i, \Ball(p_{i}, 10^{i})) \geq 10^{i}/5$,} \\
        \arg\min_{p\in \Ball(p_i, 10^i)} \cost(p, \Ball(p_i, 10^{i}))
            \opt{arxiv}{&}\opt{soda}{\\\quad}
            \mbox{otherwise.}
        \end{cases}
    \end{align*}
    \item Output $(p_0, p_1, \ldots, p_t)$.
\end{itemize}
By definition, we have that $(p_0, p_1, \ldots, p_t)$ is a $t$-robust tuple with $p=p_t$. We proceed to prove basic properties of $t$-robust tuples that will allow us to analyze our main algorithm that will repeatedly make calls to \makeHOAlg.

We start with a simple observation that bounds the distance from $p_0$ to the other points. 
\begin{lemma}
    Let $(p_0, p_1, \ldots, p_t)$ be a $t$-robust tuple and let $B_j = \Ball(p_j, 10^j)$. For every $j=1, \ldots, t$,
    \begin{gather*}
       \dist(p_{j-1}, p_j) \leq 10^{j}/2,  \qquad B_{j-1}\subseteq B_j,
       \opt{arxiv}{\qquad \mbox{and} \qquad}\opt{soda}{\mbox{ and}\\}
       \dist(p_0, p_j) \leq {10^{j}}/2\,.    
    \end{gather*}
    \label{lem:nested_balls}
\end{lemma}
\begin{proof}
    By the definition of a $t$-robust tuple,  we have $\dist(p_{j-1}, p_{j}) = 0$ if $\avgcost(p_{j}, B_j) \geq 10^{j}/5$. Otherwise, by the selection of $p_{j-1}$,  $\cost(p_{j-1}, B_j) \leq \cost(p_{j}, B_j)$ and so
    \begin{align*}
         \frac{10^{j}}{5} + \frac{10^{j}}{5} &\geq \avgcost(p_{j}, B_j) +\avgcost(p_{j-1}, B_j)\\
        & = \sum_{p\in B_j} \frac{w(p)}{w(B_j)} \left( \dist(p, p_j) + \dist(p, p_{j-1}) \right) \\
        & \geq \dist(p_j, p_{j-1}) \sum_{p\in B_j} \frac{w(p)}{w(B_j)}
    \end{align*}
    Hence, we have $\dist(p_{j-1}, p_{j}) \leq \frac{2}{5} \cdot 10^{j} \leq 10^j/2$ for all $j=1,2,\ldots, t$. It follows that any point $q\in B_{j-1}$ is also in $B_j$ since by the triangle inequality  $\dist(q, p_j) \leq \dist(q, p_{j-1}) + \dist(p_{j-1}, p_j) \leq 10^{j-1} + 10^j/2 \leq 10^j$. Hence $B_{j-1} \subseteq B_j$. Finally,  we again use   the triangle inequality to  conclude 
    \begin{align*}
        \dist(p_0, p_j)
        \leq \sum_{i=1}^{j}  \dist(p_{i-1}, p_{i})
        \opt{arxiv}{\leq \frac{2}{5} \sum_{i=1}^{j} 10^{i} = \frac{2}{5} \frac{10^{j+1}-2}{9}}
        \leq 10^j/2\,.
    \end{align*}
\end{proof}

The following lemma will be used to bound the cost of the center $p_0$ of a $t$-robust tuple in our analysis and it intuitively says that $p_0$ is a ``good'' center for any subset of $P$.

\begin{lemma}
    Let  $(p_0, p_1, \ldots, p_t)$ be a $t$-robust tuple. For $i\in \{0, 1, \ldots, t\}$ and a subset $P' \subseteq P$ of the points that contains $\Ball(p_i, 10^i)$, i.e., $\Ball(p_i, 10^i) \subseteq P'$, we have
    \begin{align*}
        \cost(p_0, P') \leq \frac{3}{2} \cdot \cost(p_i, P')\,.
    \end{align*}
    \label{lem:robust_cost}
\end{lemma}

\begin{proof}
For brevity, let $B_j= \Ball(p_j, 10^j)$ for $j= 1,2,\ldots, t$. As $(p_0, p_1, \ldots, p_t)$ is $t$-robust, we have
\begin{align}
    \cost(p_{j}, B_{j}) \geq \cost(p_{j-1}, B_{j}) \opt{arxiv}{\qquad}\opt{soda}{\quad} \mbox{for $j=1,\ldots, t$}
    \label{eq:robustcost1}
\end{align}
Furthermore, a point $q\in P \setminus B_{j}$ has distance at least $10^{j}$ from $p_{j}$ and we  have $\dist(p_0, p_j)\leq 10^j/2$ by Lemma~\ref{lem:nested_balls}.
Therefore, (by the triangle inequality)
\begin{align}
    w(q) \dist(p_{j}, q) \geq \frac{2}{3} w(q) \dist(p_0, q) \opt{arxiv}{\qquad}\opt{soda}{\quad} \mbox{for any $q\in P \setminus B_{j}$.} 
    \label{eq:robustcost2}
\end{align}
Repeatedly applying the Inequalities~\eqref{eq:robustcost1} and~\eqref{eq:robustcost2} and using $B_0 \subseteq B_1 \subseteq \cdots \subseteq B_i \subseteq P'$ (by \cref{lem:nested_balls}) give us
\begin{align*}
    \opt{soda}{& \hphantom{={}}}
    \cost(p_i, P')
    \opt{soda}{\\}\opt{arxiv}{&}
    \opt{soda}{&} = \cost(p_i, B_i) + \cost(p_i, P' \setminus B_i) \\
    & \geq \cost(p_{i-1}, B_i) + \frac{2}{3}\cost(p_0, P' \setminus B_i) \\
    \opt{arxiv}{
        & = \cost(p_{i-1}, B_{i-1}) + \cost(p_{i-1}, B_{i} \setminus B_{i-1}) + \frac{2}{3} \cost(p_0, P' \setminus B_i) \\
        & \geq \cost(p_{i-2}, B_{i-1}) + \frac{2}{3}\cost(p_0, B_{i} \setminus B_{i-1}) + \frac{2}{3} \cost(p_0, P' \setminus B_i) \\
        & = \cost(p_{i-2}, B_{i-1}) + \frac{2}{3}\cost(p_0,  P' \setminus B_{i-1}) \\
    }
    & \vdots \\
    & \geq \cost(p_0, B_1) + \frac{2}{3} \cost(p_0,  P' \setminus B_{1})  %
\end{align*}

\end{proof}

\section{Description of \epochalgo}\label{sec:epoch}
In this section we describe \epochalgo, the main new algorithm in our approach.  We assume that 
\begin{itemize}
    \item the stream is compressed into $\widetilde{O}(k)$ insertions of weighted points; and
    \item the insertions change the value of an optimum clustering by at most a factor of $18$.
\end{itemize}
These assumptions are without loss of generality and follows from the application of known techniques as described in the overview. Indeed, by losing at most a constant-factor in the approximation guarantee and poly-logarithmic factors in the consistency,  we can achieve the first assumption by an adaptation of Meyerson's sketch (similar to what was previously used for this problem in~\cite{lattanzi2017consistent}); and we can achieve the second assumption by restarting the algorithm every time the value of an optimum clustering increases by a constant-factor (which can happen at most $O(\log(n\Delta)$ times). For a more formal description of this reduction please refer to \cref{sec:everything_alg}.

The compressed stream is now divided into epochs. In each epoch we call \epochalgo which takes as input the final clustering produced during the last epoch (or an initial solution if we consider the first epoch).  In order to guarantee 
a constant approximate solution along with an amortized poly-logarithmic number of changes in the solution, we require that the set of input centers in each call to \epochalgo has the following properties.
\begin{Definition} \label{def:req-prop}
We call a set of centers $\mathcal{U}$ {\emph{bounded-robust}} if it has the following two properties.
\begin{enumerate}
    \item $\mathcal{U}$ is a $100$-approximate solution.
    \item Each center $u\in \mathcal{U}$ is $t$-robust where $t$ is the smallest integer such that $10^t \geq \dist(u, \mathcal{U}\setminus \{u\})/200$.
\end{enumerate}
\end{Definition}

In the first epoch, we construct an initial bounded-robust solution  to the point set  as follows:  we first obtain a $10$-approximate solution $\mathcal{W}$ to the initial point set using one of the known constant-factor approximation algorithms for $k$-median; and  we then \emph{robustify} the centers of $\mathcal{W}$ as follows: \\[1mm]
\vspace{-4mm}
    \begin{center}
\begin{minipage}{0.95\columnwidth}
\begin{mdframed}[hidealllines=true, backgroundcolor=gray!15]
    While there is a center $w\in \mathcal{W}$ that violates the second condition of bounded-robust, i.e., it is not $t$-robust for the smallest integer $t$ such that $10^t \geq \dist(w, \mathcal{W} \setminus \{w\})/200$: %
    \begin{enumerate}
        \item Obtain $t'$-robust tuple $(w_0, w_1, \ldots, w_{t'})  = \makeHOAlg(t', w)$ with $w_{t'} = w$  for smallest  integer $t'$ such that $10^{t'} \geq \dist(w, \mathcal{W} \setminus w)/100$.
        
        \item Remove $w$ and add $w_0$ to the set of cluster centers.
    \end{enumerate}
\end{mdframed}
\end{minipage}
\end{center}
We refer to the above procedure that makes every center in $\mathcal{W}$ robust as $\algrobustify(\mathcal{W})$.
We prove that \makeHOAlg is called at most once for each center (see \cref{lem:makerobustatmostonce}) and so the above procedure terminates with a set of cluster centers that satisfies the second condition of \cref{def:req-prop}. For the analysis, we may actually assume that \makeHOAlg is also called at least (i.e., exactly) once for each center. That the procedure satisfies the first condition follows from the fact that the cost is only increased by a factor $3/2$ by robustifying the cluster centers in $\mathcal{W}$ (see \cref{lemma:cost-any-robust}).

Let $\mathcal{U}^{(0)}$  denote the  bounded-robust cluster centers obtained for the initial point set $P^{(0)}$.   \epochalgo  first detects the number of centers, $\ell$, that can be removed from the current solution $\mathcal{U}^{(0)}$ without affecting the quality of the approximation too much; then it handles $\ell+1$ insertions, i.e., produces $\ell+1$ solutions $U^{(1)}, U^{(2)}, \ldots, U^{(\ell+1)}$ for point sets $P^{(1)}, P^{(2)}, \ldots, P^{(\ell+1)}$, where $P^{(i)}$ denotes the set of points obtained from $P^{(0)}$ after $i$ weighted insertions. The last step ensures that the set of centers $\mathcal{U}^{(\ell+1)}$ is bounded-robust and is thus a valid initial solution in the next epoch where \epochalgo is called with $U^{(\ell+1)}$ as the ``$\mathcal{U}^{(0)}$''-solution and $P^{(\ell+1)}$ as the ``$P^{(0)}$'' point set.  \epochalgo is repeatedly called in this way until all insertions are considered in the compressed stream. 

We proceed to describe the steps of \epochalgo in detail when given as  input  a set $\mathcal{U}^{(0)}$ of bounded-robust centers  of points $P^{(0)}$:

\paragraph{Step 1: Removing Centers.} In this preprocessing step, we remove all the centers that are not necessary in our solution. %
Basically we remove $\ell$ centers while increasing the cost of the solution by a factor at most $O(1)$. More precisely, if we let $\ell^*$ be the largest value such that it is possible to remove $\ell^*$ centers from $\mathcal{U}^{(0)}$ while increasing the cost by at most a factor $c = 228000$. Then we find a solution of size at most $k-\ell$ with $\ell \geq \ell^*$ and cost at most $3 c \cdot \cost(\mathcal{U}^{(0)}, P^{(0)})$.   The details of this step is described in \cref{sec:removing-centers} (\cref{thm:lpremove}) . In short, we use the standard LP to find the value of $\ell$ and then we apply  known algorithmic techniques for the $k$-median problem. 

\paragraph{Step 2: Handling Insertions.} From the preprocessing step, we know that the size of the current solution is at most $k- \ell$. This enables us to simply open the next $\ell$ point that are inserted as centers, so the cost of the solution does not increase. We refer to the clustering and the set of points after the $i$-th insertion as $\mathcal{U}^{(i)}$ and $P^{(i)}$ for $1 \leq i \leq \ell$, respectively. Then, we consider one more insertion, i.e., the point set $P^{(\ell+1)}$.  After this insertion the cost of the solution may increase significantly and furthermore the current solution may not be $t$-robust. We address both issues in the next step.

\paragraph{Step 3: Swapping Center and  Robustify.} In this postprocessing step, the first goal is to find a set of $O(\ell+1)$ swaps that minimizes the objective function. This enables us to bound the approximation ratio of our approach and prove that it is a $100$-approximate solution. To that end,  for $\ell' = 5\ell+ 5$, we use an LP-rounding procedure that swaps $4\ell'$ elements from $\mathcal{U}^{(0)}$ and produces $\mathcal{W}$. We show that $\mathcal{W}$ is a $13$-approximate solution with respect to the cost of an optimum solution $\mathcal{W}_{\textrm{OPT}}$ that swaps at most $\ell'$ centers, with probability at least $1-(n+\Delta)^{10}$.  The details of this procedure is described in \cref{sec:swapping-centers} (\cref{thm:lpswap}).

Finally, we make the solution bounded-robust so that it can be used in the next \epochalgo using \makeHOAlg.
More precisely, we robustify the cluster centers in $\mathcal{W}$ to obtain $U^{(\ell+1)}$ by calling $\algrobustify(\mathcal{W})$  as we did for the initial solution (described in the gray box above).  

\paragraph{}
Having described our main algorithm we proceed to its analysis. In the next section, we first bound the number of changes to the solution we make, i.e., the consistency. Then in \cref{sec:cost-of-epoch} we bound the cost of the maintained solution. In particular, we show that $\mathcal{U}^{(\ell+1)}$ is indeed a bounded-robust solution and thus a valid input to the next call to \epochalgo. 
\section{Bound on the Number of Changes}\label{sec:robustifications}
We analyze the consistency, i.e., the number of changes made to the maintained solution. As we will see,  the number of changes essentially boils down to analyzing the number  of calls to \makeHOAlg in \algrobustify. We first observe that \algrobustify calls \makeHOAlg at most once for each center. This implies that the procedure terminates and it is also a fact that will be used in the cost analysis. We then proceed to the main part of this section, which is the consistency analysis. 

\subsection{Robustify Makes a Center Robust At Most Once}
We start by showing that \algrobustify calls \makeHOAlg at most once for each center. This guarantees that \algrobustify  terminates and it will later also be used to  bound the cost of the clustering.
\begin{lemma}
    Consider a set $\mathcal{W}$ of centers. If $\algrobustify(\mathcal{W})$ calls \makeHOAlg for center $w\in W$ and thus replaces $w$ by $w_0$, then it makes no subsequent call to \makeHOAlg for center $w_0$. 
    \label{lem:makerobustatmostonce}
\end{lemma}
\begin{proof}
    The statement follows intuitively due to the following. When a call to $\makeHOAlg$ on center $w$ is made, it is with the smallest $t$ such that $10^t \geq \dist(w, \mathcal{W}\setminus \{w\})/100$. Center $w$ is then replaced by a nearby center $w_0$ and, for $w_0$ to be selected in a subsequent iteration, it must be that at that point $\dist(w_0, \mathcal{W}\setminus \{w_0\})/200 > 10^t$, i.e., other centers are now more than a factor $2$ further away from $w_0$ than they were for $w$. This cannot happen since calls to \makeHOAlg only changes the position of  centers relatively  little, see \cref{lem:nested_balls}. 
    
    We proceed with the formal proof. 
    Suppose toward contradiction that there is a center $w\in \mathcal{W}$ such that $\algrobustify(\mathcal{W})$ calls  \makeHOAlg for  center $w$, which is replaced by $w_0$, and then in a subsequent iteration makes a call to \makeHOAlg for center $w_0$.
    
    We consider the first pair $w, w_0$ for which this happens. %
    Let $\mathcal{W}'$ be the set of centers when \makeHOAlg is called for $w$ and let $w' \in \mathcal{W}' \setminus \{w\}$ be a center such that $\dist(w, \mathcal{W}' \setminus \{w\}) = \dist(w, w')$. 
    Then \algrobustify calls $\makeHOAlg(t, w)$ where $t$ is selected to be the smallest integer such that $10^{t} \geq \dist(w,w')/100$. 
    In particular, we have $10^{t} \leq \dist(w,w')/10$ and so by \cref{lem:nested_balls} 
    \begin{align*}
        \dist(w_0, w) \leq 10^{t}/2 \leq \dist(w,w')/20\,.
    \end{align*}
    
    Now let $\mathcal{W}''$ be the set of centers when \makeHOAlg is called for $w_0$, then we have that either $w'\in \mathcal{W}''$ or $\mathcal{W}''$ contains the center $w'_0$ that replaced $w'$ via a single  call to \makeHOAlg. These two cases are exhaustive since $w$ and $w_0$ was the first pair such that an additional call for $w_0$ was made. 
    In the first case,
    \begin{align*}
        \dist(w_0, \mathcal{W}'') \leq \dist(w_0, w') \opt{soda}{&} \leq \dist(w_0, w) + \dist(w, w') \opt{soda}{\\ &} \leq \frac{21}{20} \dist(w,w')
    \end{align*}
    which contradicts that $w_0$ was selected in the while-loop in \algrobustify since we have that $w_0$ is $t$-robust with $10^{t}\geq \dist(w,w')/100$ and thus $10^{t} \geq \dist(w_0, \mathcal{W}'')/200$.

    The second case is similar but we need in addition to argue that $\dist(w'_0, w') \leq \dist(w, w')/10$. To see this note that \algrobustify calls $\makeHOAlg(t', w')$ with a $t'$ such that $10^{t'} \leq \dist(w', w_0)/10$. Hence, \cref{lem:nested_balls} says that $\dist(w'_0, w') \leq 10^{t'}/2 \leq \dist(w', w_0)/20$ which in turn is upper bounded by $(\dist(w_0,w)+ \dist(w,w'))/20 \leq \dist(w,w')/10$. We thus have in the second case that
    \begin{align*}
        \opt{soda}{& \hphantom{={}}} \dist(w_0, \mathcal{W}'') \opt{soda}{\\ &} \leq \dist(w_0, w'_0) \opt{soda}{\\} & \leq \dist(w_0, w) + \dist(w, w') + \dist(w', w'_0)  \\
        & \leq \dist(w,w')/20 + \dist(w,w') + \dist(w, w')/10 \\
        & = \frac{23}{20} \dist(w,w')\,,
    \end{align*}
     which contradicts that $w_0$ was selected in the while-loop in \algrobustify in the same way as in the previous case. 
    
\end{proof}

\subsection{Consistency Analysis}
In this section we focus on analyzing the consistency of the algorithm. Recall that $m= \widetilde{O}(k)$ denotes the number of weighted insertions in our compressed stream.
\begin{theorem} \label{thm:epoc-consistency}
    The number of changes to the maintained solution is at most $O(m (\log \aspratio)^2)$.
    \label{thm:consistency}
\end{theorem}

We use the following notation.   Let $E$ denote the total number of epochs. We use the convention that we subscript quantities in the call of \epochalgo during the $e$-th epoch by $e$. So,  for $1\leq e\leq E$, the set of centers $\mathcal{U}^{(0)}_e$ denotes the initial solution to point set $P^{(0)}_e$ in the call to \epochalgo at the start of the  $e$-th epoch, and $\mathcal{U}^{(1)}_e, \mathcal{U}^{(2)}_e, \ldots, \mathcal{U}^{(\ell_e+1)}_e$  denote the solutions produced to point sets $P^{(1)}_e, P^{(2)}_e, \ldots, P^{(\ell_e+1)}_e$ during this run of \epochalgo.  With this notation, we have $U^{(\ell_e +1)}_e = U^{(0)}_{e+1}$ and $P^{(\ell_e+1)}_e = P^{(0)}_{e+1}$ for $e = 1, 2,\ldots, E-1$. Moreover, the consistency equals
\begin{align}
    \sum_{e=1}^E \sum_{i=0}^{\ell_e}|\mathcal{U}^{(i+1)}_e \setminus \mathcal{U}^{(i)}_e|\,.
    \label{eq:consistency}
\end{align}

The consistency $\sum_{i=0}^{\ell_e} |\mathcal{U}^{(i+1)}_e \setminus \mathcal{U}^{(i)}_e|$ of epoch $e$ is  at most $\ell_e + |\mathcal{U}_e^{(\ell_e+1)} \setminus \mathcal{U}_e^{(\ell_e)}|$ since \epochalgo opens up exactly one center for $i=1,2,\ldots, \ell$.
We further have
\begin{align*}
    |\mathcal{U}_e^{(\ell_e+1)} \setminus \mathcal{U}_e^{(\ell_e)}| \opt{soda}{&} \leq 
    |\mathcal{U}_e^{(\ell_e+1)} \setminus \mathcal{U}_e^{(0)}| + |\mathcal{U}_e^{(0)} \setminus \mathcal{U}_e^{(\ell_e)}| \opt{soda}{\\}
    \opt{soda}{&} = |\mathcal{U}_e^{(\ell_e+1)} \setminus \mathcal{U}_e^{(0)}| + O(\ell_e)\,. 
\end{align*}
This allows us to upper bound \cref{eq:consistency} by
\begin{align*}
    \opt{soda}{&} \sum_{e=1}^E \left( |\mathcal{U}_e^{(\ell_e+1)} \setminus \mathcal{U}_e^{(0)}| + O(\ell_e) \right) \opt{soda}{\\}
    \leq \opt{soda}{&} \sum_{e=1}^E |\mathcal{U}_e^{(\ell_e+1)} \setminus \mathcal{U}_e^{(0)}| + O(m)\,,
\end{align*}
where we used that $\sum_{e=1}^E (\ell_e +1) = m$.

To analyze $|\mathcal{U}_e^{(\ell_e+1)} \setminus \mathcal{U}_e^{(0)}|$, recall that \epochalgo constructs $\mathcal{U}_e^{(\ell_e+1)}$ in two steps: it first obtains an intermediate solution $\mathcal{W}_e$ by swapping $4\ell'_e = 20(\ell_e+1)$ centers from $\mathcal{U}_e^{(0)}$; and it then calls $\algrobustify(\mathcal{W}_e)$ to obtain $\mathcal{U}_e^{(\ell_e+1)}$.

We distinguish two kind of centers in $\mathcal{U}_e^{(\ell_e +1)} \setminus \mathcal{U}_e^{(0)}$, the updated centers and the new centers.  To this end, 
consider a center $w\in \mathcal{W}_e \cap \mathcal{U}^{(0)}_e$. If $w \not \in \mathcal{U}^{(\ell_e+1)}_e$, then $w$ was replaced by a new center $w_0$ via  call to \makeHOAlg in the \algrobustify procedure. We will say that $w$ is the \emph{parent} of $w_0$ and we say that $w_0$ is an \emph{updated} center. Those centers in $U^{(\ell_e+1)}_e \setminus U^{(0)}_e$ that are not updated (i.e., without a parent) are referred to as \emph{new} centers.  In addition the centers in the initial solution computer for the first epoch, are also classified as new centers.

It will be convenient to think of these centers as elements of chains formed as follows.
Consider the graph that has a vertex for each center in $\mathcal{U}_e^{(\ell_e+1)} \setminus \mathcal{U}_e^{(0)}$ for each epoch $1\leq e\leq E$ and there is an arc from $u$ to $u_0$ if $u$ is the parent of $u_0$. We remark that the graph may have multiple vertices for a single center, if that center was added and removed from the solution multiple times.  As each vertex has at most one parent and it is the parent of at most one center, we have the graph forms a collection of paths.
We refer to these paths as chains. Moreover, if we consider a chain $(u_1, u_2, \ldots, u_s)$ then $u_1$ is a new center (without a parent), $u_{i+1}$ replaced $u_i$ via a call to \makeHOAlg for $i=1,2,\ldots, s-1$, and $u_s$ is not the parent of any center.

By definition, we have that 
$\sum_{e=1}^E |\mathcal{U}_e^{(\ell_e+1)} \setminus \mathcal{U}_e^{(0)}|$ equals the number of vertices in the above-described graph. Or equivalently, and this is the viewpoint that we take,  it equals the total lengths of the chains. We start to bound the number of chains.
\begin{lemma}
    The total number of chains (or equivalently, new centers) is at most $20\cdot m$.
    \label{lem:newcenters}
\end{lemma}
\begin{proof}
  Consider an epoch $e$. We have that a center in $\mathcal{W}_e\cap \mathcal{U}^{(0)}_e$ is either in $\mathcal{U}^{(\ell+1)}_e$ or it is the parent of an updated center in $\mathcal{U}^{(\ell+1)}$. It follows that the number of new centers in $\mathcal{U}_e^{(\ell_e+1)} \setminus \mathcal{U}_e^{(0)}$ is at most $|\mathcal{U}_e^{(\ell_e+1)}| - |\mathcal{W}_e \cap \mathcal{U}^{(0)}_e|$. Now $\mathcal{W}_e$ is obtained from $\mathcal{U}^{(0)}_e$ by performing at most $4\ell'_e = 20(\ell_e+1)$ swaps and so  $|\mathcal{W}_e \cap \mathcal{U}^{(0)}_e| \geq k - 20(\ell_e +1)$. Hence, we can upper bound the number of new centers in $\mathcal{U}_e^{(\ell_e+1)} \setminus \mathcal{U}_e^{(0)}$ by $20(\ell_e+1)$. It follows that the number of new centers across all epochs is upper bounded by $20\sum_{e=1}^E (\ell_e+1) = 20\cdot m$. Note that chains can start only in new centers by the definition of new centers.
\end{proof}

We now give certain properties that must hold when \algrobustify calls \makeHOAlg for a center $w$ and then replaces it.
 To this end, we use the following corollary   of \cref{def:hierarchically_optimal,lem:nested_balls}, to identify points that get inserted and invalidate the robustness guarantees maintained by the algorithm.
\begin{corollary}
    \label{cor:certificate-invalidators}
    Consider a point set $P$, and suppose that the center $u\in P$ is $t$-robust. 
    Then $u$ is $t$-robust for every superset $P' \supseteq P$ of points satisfying  $\Ball_P(u, 2\cdot 10^{t}) = \Ball_{P'}(u, 2\cdot 10^{t})$, i.e., $P' \setminus P$ does not contain any point within distance $2\cdot 10^{t}$ from $u$.  
\end{corollary}
\begin{proof}
    Let $p_0 = u$ and let $(p_0, p_1, \ldots, p_t)$ be a $t$-robust sequence with respect to $P$. Then 
    \cref{def:hierarchically_optimal,lem:nested_balls} imply that $(p_0, p_1, \ldots, p_t)$ is also a $t$-robust sequence with respect to $P'$ if $\Ball_P(p_t,  10^{t}) = \Ball_{P'}(p_t,  10^{t})$. The statement now follows from \cref{lem:nested_balls} which says that $\dist(p_0, p_t) \leq 10^{t}/2$ and thus any point in a ball of radius $10^t$ around $p_t$ is contained in a ball of radius $2\cdot 10^t$ around $p_0 = u$.
\end{proof}

For a center $u$ that is in one of the maintained solutions, define the integer $t(u)$ as follows. 
If $u$ is a new center, let $t(u) = 0$. Otherwise  if $u$ is an updated center, then we let $t(u)$ be the integer $t$ used in the call to $\makeHOAlg(t,u')$ by \algrobustify when it replaced $u$'s parent $u'$ by $u$.

Now consider an updated center $u_0$ with parent $u$. Let $f$ be the epoch when $u$ joined the solution for the last time before $u_0$ replaced $u$ and let $e$ be the epoch when it was replaced. %
As \algrobustify replaced $u$ with $u_0$, at least one of the following two cases must hold: 
\begin{itemize}
    \item Center $u$ is not $t(u)$-robust with respect to the bigger point sets $P^{(\ell_e+1)}_e$. By the above corollary, there must then be a point in  $p\in P^{(\ell_e+1)}_e \setminus  P^{(\ell_f+1)}_f$ within distance $2\cdot 10^{t(u)}$ from $u$. We say that the point $p$ invalidates the center $u$ in this case.
    \item  The integer $t(u)$ is too small: When $w=u$ is selected in the while-loop of \algrobustify, we have $10^{t(w)}< \dist(w, \mathcal{W}\setminus \{w\})/200$. Then  $u_0$ is obtained via a call to $\makeHOAlg(t', u)$ where  $t'$ is the smallest integer such that $10^{t'}> \dist(w, \mathcal{W}\setminus \{w\})/100$. In particular, $t(u_0) = t'\geq t+1$. In this case, we say that the updated center $u_0$ \emph{increased its robustness}.
\end{itemize}

We first bound the number of updates that increases the robustness in terms of the number of new centers and other updated centers. We then bound the number of updated centers that were invalidated by points.

\begin{lemma}
    The number of updated centers that increased their robustness is at most $O(\log \aspratio)$ times the number of new centers plus the number of updated centers that were invalidated by points.
    \label{lem:increasrobustness}
\end{lemma}
\begin{proof}
    Consider a chain $(u_0, u_1, \ldots, u_s)$. We have that $u_0$ is a new center.
    The statement now follows by observing that we cannot have a subsequence of updated centers $u_i, u_{i+1}, \ldots, u_j$ so that  everyone increases their robustness   $j- i> \log \aspratio$. Indeed, in that case, we have $t(u_j) \geq j-i > \log \aspratio$ which is a contradiction because for any center $u$ we have that $10^{t(u)}$ is at most the maximum distance $\Delta$.  Therefore a chain of length $s$ must include at least $\lfloor s/\log(\aspratio)\rfloor$ centers that are invalidated by points and for which the robustness is not increased.
\end{proof}

\begin{lemma}
    The number of updated centers that were invalidated by points is at most $m \cdot \log \aspratio$.
    \label{lem:invalidated}
\end{lemma}
\begin{proof}
    The stream consists of $m$ point insertions so it is sufficient to prove that each point $p$ invalidates at most $\log \aspratio$ many updated centers. We bound the number of centers that are invalidated by a point $p$ introduced in epoch $e$, i.e., $p\in P^{(\ell+1)}_e \setminus P^{(\ell+1)}_{e-1}$. Let $\mu$ be the number of centers in $\mathcal{W}_e\cap \mathcal{U}^{(0)}_e$ that are invalidated by $p$, and let $(u_i)_{i \in \mu}$ be the sequence of these centers ordered in decreasing order by the epoch that they were added to the maintained solution: letting $e_i$ equal the first epoch when $u_i \in \mathcal{U}^{(\ell+1)}_{e_i}$, we order the sequence so that $e_i > e_{i+1}$ for $i=1,2, \ldots, \mu-1$.
    
    This ordering guarantees that when $u_i$ replaced its parent $u'_i$ via a call to $\makeHOAlg$ then $u_{i+1}$ was already in the solution and so 
    \begin{align*}
    10^{t(u_i)} \leq \dist(u_i, u_{i+1})/10\qquad \mbox{for $i\in \{1,2,\ldots, \mu-1\}$.}
    \end{align*}
    In addition, as $p$ invalidates $u_i$, we have
    \begin{align*}
        \dist(p, u_i) \leq 2\cdot 10^{t(u_i)} \qquad \mbox{for $i\in \{1,2, \ldots, \mu\}$.}
    \end{align*}
    It follows that for $i\in \{1,2,\ldots, \mu-1\}$
    \begin{align*}
      10^{t(u_{i+1})} \opt{soda}{&} \geq \dist(p, u_{i+1}) \geq \dist(u_i, u_{i+1}) - \dist(p, u_i) \opt{soda}{\\&} \geq 10 \cdot 10^{t(u_i)} - 2\cdot 10^{t(u_i)} \geq 8 \cdot 10^{t(u_i)}\,.
    \end{align*}
    We can now conclude the proof by observing that due to the bounded aspect ratio $\aspratio$, $\mu \leq \log \aspratio$. 
\end{proof}

Equipped with the above lemmas, we can complete the proof of \cref{thm:consistency}. \cref{lem:newcenters} says that there are $O(m)$ new centers and \cref{lem:invalidated} says that there are at most $O(m \log \aspratio)$ number of updated centers that were invalidated by points. Combining this with \cref{lem:increasrobustness} yields that the total number of new and updated centers is at most $O(m (\log \aspratio)^2)$. As the number of new and updated centers is equal to the consistency, this completes the proof of the theorem.

\section{Cost Analysis of \epochalgo}
\label{sec:cost-of-epoch}

Recall that \epochalgo runs during $\ell+1$ steps and and maintains sets of centers $\mathcal{U}^{(0)}, \mathcal{U}^{(1)}, \ldots, \mathcal{U}^{(\ell+1)}$ for the point sets $P^{(0)}, P^{(1)}, \ldots, P^{(\ell+1)}$. We bound their costs as follows, recall that $c = 228000$.
\begin{theorem} \label{main-epoc-cost}
    On input a bounded-robust set of centers $\mathcal{U}^{(0)}$ of point set $P^{(0)}$, \epochalgo produces  sets of centers $\mathcal{U}^{(1)}, \mathcal{U}^{(2)}, \ldots, \mathcal{U}^{(\ell+1)}$ of the point sets $P^{(0)}, P^{(1)}, \ldots, P^{(\ell+1)}$ satisfying:
    \begin{itemize}
        \item the clustering induced by $\mathcal{U}^{(i)}$ is a $\left(6 \cdot 100\cdot c\right)$-approximate clustering for $P^{(i)}$ for $i=1,2,\ldots, \ell$; and
        \item $\mathcal{U}^{(\ell+1)}$ is a bounded-robust set of centers for $P^{(\ell+1)}$.
    \end{itemize}
    \label{thm:maincost}
\end{theorem}

The first part of the theorem follows easily: As $\mathcal{U}^{(0)}$ is a bounded-robust set of centers of $P^{(0)}$ it is a $100$-approximation. Now \epochalgo drops $\ell$ centers from $\mathcal{U}^{(0)}$ while increasing the cost by  at most a factor $3 \cdot c$. The cost of the solutions $\mathcal{U}^{(1)}, \mathcal{U}^{(2)}, \ldots, \mathcal{U}^{(\ell)}$ does not change (since the newly arrived points are opened as centers) and thus they remain $(2 \cdot 3 \cdot 100 \cdot c)$-approximate solutions. The factor $2$ is due to the fact that the value of an optimum solution can at most go down by a factor of $2$ by introducing new points. 

The remaining part of this section is thus devoted to proving the second part of the theorem, i.e.,  that $U^{(\ell+1)}$ is a bounded-robust set of centers for $P^{(\ell+1)}$.  
Recall that Step 3 of \epochalgo calls \algrobustify($\mathcal{W}$) procedure which repeatedly call \makeHOAlg until every center in $u \in \mathcal{W}$ is $t$-robust with $10^t \geq \dist\left(u, \mathcal{W}\setminus \{u\}\right)/100$. Therefore it suffice to showing that $\mathcal{U}^{(\ell+1)}$ induce a $100$-approximate clustering of $P^{(\ell+1)}$.

An important concept for bounding the cost of $\mathcal{U}^{(\ell+1)}$ is the notion of well-separated pairs. To simplify notation, we let $\mathcal{U}$ denote $\mathcal{U}^{(0)}$ throughout this section.  We also let $\mathcal{V}$ be a fixed optimal solution to $P^{(\ell+1)}$.
\begin{Definition}\label{def:well-separated}
    Select $\gamma = 2000$. We say that centers $u\in \mathcal{U}, v\in \mathcal{V}$ form a \emph{well-separated pair} if %
    \begin{gather*}
        \dist\left(u, \mathcal{U} \setminus \{u\}\right) \geq \gamma \cdot   \dist(u,v) \opt{soda}{\\} \opt{arxiv}{\qquad} \mbox{ and } \opt{arxiv}{\qquad} \dist\left(v, \mathcal{V} \setminus \{v\}\right) \geq \gamma \cdot   \dist(u,v)\,.
    \end{gather*}
\end{Definition}
\begin{figure}[h]
    \centering
    \subfloat[]{%
        \label{fig:ca-well-separated}%
        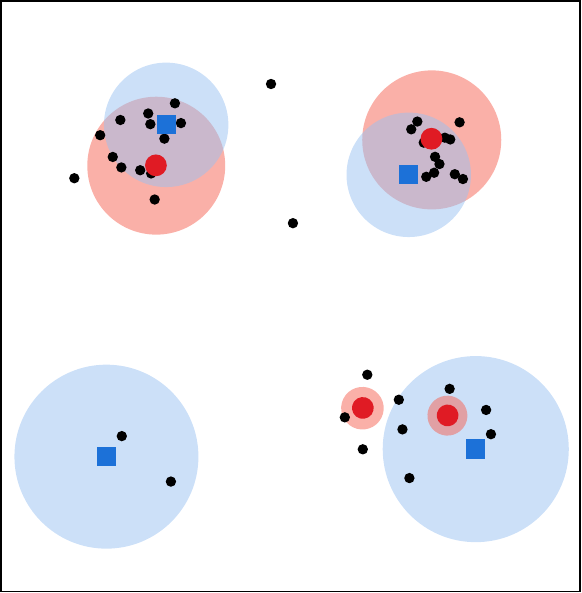%
    }%
    \qquad%
    \subfloat[]{%
        \label{fig:ca-lp2}%
\begingroup%
  \makeatletter%
  \providecommand\color[2][]{%
    \errmessage{(Inkscape) Color is used for the text in Inkscape, but the package 'color.sty' is not loaded}%
    \renewcommand\color[2][]{}%
  }%
  \providecommand\transparent[1]{%
    \errmessage{(Inkscape) Transparency is used (non-zero) for the text in Inkscape, but the package 'transparent.sty' is not loaded}%
    \renewcommand\transparent[1]{}%
  }%
  \providecommand\rotatebox[2]{#2}%
  \newcommand*\fsize{\dimexpr\f@size pt\relax}%
  \newcommand*\lineheight[1]{\fontsize{\fsize}{#1\fsize}\selectfont}%
  \ifx\svgwidth\undefined%
    \setlength{\unitlength}{85.59540155bp}%
    \ifx\svgscale\undefined%
      \relax%
    \else%
      \setlength{\unitlength}{\unitlength * \real{\svgscale}}%
    \fi%
  \else%
    \setlength{\unitlength}{\svgwidth}%
  \fi%
  \global\let\svgwidth\undefined%
  \global\let\svgscale\undefined%
  \makeatother%
  \begin{picture}(1,1.99284931)%
    \lineheight{1}%
    \setlength\tabcolsep{0pt}%
    \put(0,0){\includegraphics[width=\unitlength,page=1]{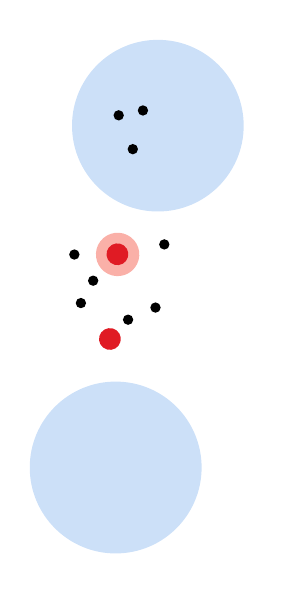}}%
    \put(0.44038589,1.02462736){\makebox(0,0)[lt]{\lineheight{1.25}\smash{\begin{tabular}[t]{l}$u_p$\end{tabular}}}}%
    \put(0,0){\includegraphics[width=\unitlength,page=2]{LPcase2.pdf}}%
    \put(0.54719124,1.45368543){\makebox(0,0)[lt]{\lineheight{1.25}\smash{\begin{tabular}[t]{l}$v_p$\end{tabular}}}}%
    \put(0,0){\includegraphics[width=\unitlength,page=3]{LPcase2.pdf}}%
    \put(0.64296705,1.138438){\makebox(0,0)[lt]{\lineheight{1.25}\smash{\begin{tabular}[t]{l}$p$\end{tabular}}}}%
    \put(0.42496112,0.7418678){\makebox(0,0)[lt]{\lineheight{1.25}\smash{\begin{tabular}[t]{l}$u'_p$\end{tabular}}}}%
    \put(0,0){\includegraphics[width=\unitlength,page=4]{LPcase2.pdf}}%
  \end{picture}%
\endgroup%
    }%
    \qquad%
    \subfloat[]{%
        \label{fig:ca-lp3}%
\begingroup%
  \makeatletter%
  \providecommand\color[2][]{%
    \errmessage{(Inkscape) Color is used for the text in Inkscape, but the package 'color.sty' is not loaded}%
    \renewcommand\color[2][]{}%
  }%
  \providecommand\transparent[1]{%
    \errmessage{(Inkscape) Transparency is used (non-zero) for the text in Inkscape, but the package 'transparent.sty' is not loaded}%
    \renewcommand\transparent[1]{}%
  }%
  \providecommand\rotatebox[2]{#2}%
  \newcommand*\fsize{\dimexpr\f@size pt\relax}%
  \newcommand*\lineheight[1]{\fontsize{\fsize}{#1\fsize}\selectfont}%
  \ifx\svgwidth\undefined%
    \setlength{\unitlength}{80.88314807bp}%
    \ifx\svgscale\undefined%
      \relax%
    \else%
      \setlength{\unitlength}{\unitlength * \real{\svgscale}}%
    \fi%
  \else%
    \setlength{\unitlength}{\svgwidth}%
  \fi%
  \global\let\svgwidth\undefined%
  \global\let\svgscale\undefined%
  \makeatother%
  \begin{picture}(1,2.10895274)%
    \lineheight{1}%
    \setlength\tabcolsep{0pt}%
    \put(0,0){\includegraphics[width=\unitlength,page=1]{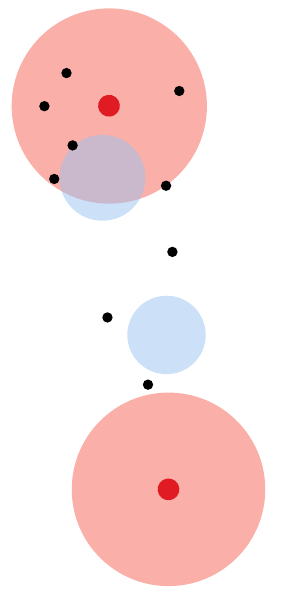}}%
    \put(0.44869795,1.63617382){\makebox(0,0)[lt]{\lineheight{1.25}\smash{\begin{tabular}[t]{l}$u_p$\end{tabular}}}}%
    \put(0,0){\includegraphics[width=\unitlength,page=2]{LPcase3.pdf}}%
    \put(0.41809337,1.32290176){\makebox(0,0)[lt]{\lineheight{1.25}\smash{\begin{tabular}[t]{l}$v_p$\end{tabular}}}}%
    \put(0,0){\includegraphics[width=\unitlength,page=3]{LPcase3.pdf}}%
    \put(0.79830711,1.76976845){\makebox(0,0)[lt]{\lineheight{1.25}\smash{\begin{tabular}[t]{l}$p$\end{tabular}}}}%
    \put(0,0){\includegraphics[width=\unitlength,page=4]{LPcase3.pdf}}%
    \put(0.64765026,0.79544843){\makebox(0,0)[lt]{\lineheight{1.25}\smash{\begin{tabular}[t]{l}$v'_p$\end{tabular}}}}%
    \put(0.63470847,0.23874321){\makebox(0,0)[lt]{\lineheight{1.25}\smash{\begin{tabular}[t]{l}$u'_p$\end{tabular}}}}%
    \put(0,0){\includegraphics[width=\unitlength,page=5]{LPcase3.pdf}}%
  \end{picture}%
\endgroup%
    }%
    \caption{%
        Each figure shows two sets of centers $\mathcal{U}$ (red circles) and $\mathcal{V}$ (blue squares). For each center $u \in \mathcal{U}$, $\Ball(u, \dist(u, \mathcal{U} \setminus \{ u \}) / \gamma)$ is depicted, and similarly for centers $v \in \mathcal{V}$ (for this figure only, we assume $\gamma = 4$). Therefore $(u,v) \in \mathcal{U} \times \mathcal{V}$ is a well-separated pair if and only if $u$ lies within the ball of $v$ \emph{and} vice versa. %
        \protect\subref{fig:ca-well-separated} The upper two pairs are well-separated, the lower centers are not well-separated. %
        \protect\subref{fig:ca-lp2} Fractional assignment: the case that $y_{u_p} = 1/2$ and there is a center $u'_p\in \mathcal{U} \setminus \{u_p\}$ such that $\dist(u_p, u'_p) \leq \gamma\cdot  \dist(u_p, v_p)$. %
        \protect\subref{fig:ca-lp3} Fractional assignment: the case that $y_{u_p} = 1/2$ and the previous case \protect\subref{fig:ca-lp2} does not hold.}
    \label{fig:cost-analsis}
\end{figure}
Informally, $u\in \mathcal{U}$ and $v\in \mathcal{V}$ form a well-separated pair if they are isolated in the sense that they are much closer to each other than the distance to any other center. In particular, by definition, if $u$ and $v$ form a well-separated pair, then $v$ must be the closest center to $u$ in $\mathcal{V}$ and $u$ must be the closest center to $v$ in $\mathcal{U}$ (see \cref{fig:ca-well-separated}). We can therefore partition the centers in $\mathcal{U} \cup \mathcal{V}$ into the well-separated pairs $(u_1, v_1), (u_2, v_2), \ldots, (u_{k-m}, v_{k-m})$ and the remaining centers $u_{k-m+1}, \ldots, u_k, v_{k-m+1}, \ldots, v_k$ that do not form a well-separated pair with any center.  In \cref{sec:number_wellseparated} we prove the following lemma which relates the number of well-separated pairs with the number of centers we can drop from $\mathcal{U}$ without increasing the cost too much.
\begin{lemma}
    \label{lem:well_separeted_remove}
   Suppose that the number of well-separated pairs is $k-m$. Then there exists a clustering $\mathcal{U}' \subseteq \mathcal{U}$ with at most $k- \lfloor m/4 \rfloor$ centers and whose cost is bounded by
   \begin{align*}
       \cost\left(\mathcal{U}', P^{(0)}\right) \leq 6\gamma \left(\cost\left(\mathcal{U}, P^{(0)}\right) + \cost\left(\mathcal{V}, P^{(0)}\right)\right)\,.
   \end{align*}
\end{lemma}

By the assumption that the value of an optimum clustering increases by at most  a factor $18$ during an epoch, we have, by the optimality of $\mathcal{V}$,  that $\cost\left(\mathcal{V}, P^{(0)}\right)$ is within a factor of $18$ of the cost of an optimal clustering of $P^{(0)}$. It follows  that  $6\gamma \left(\cost\left(\mathcal{U}, P^{(0)}\right) + \cost\left(\mathcal{V}, P^{(0)}\right)\right) \leq (6\gamma \cdot  (18+1)) \cost\left(\mathcal{U}, P^{(0)}\right)$. As $c = 228000 = 6 \gamma \cdot 19$, this implies that Step~$1$ of \epochalgo removes $\ell\geq \lfloor m/4 \rfloor$ centers where $m$ is the number of centers of $\mathcal{U}$ that does not form a well-separated with a center in $\mathcal{V}$. 

During Step~3, \epochalgo  creates an intermediate clustering $\mathcal{W}$ whose cost is $13$-approximate with  respect to the cost of an optimum solution $\mathcal{W}_{\textrm{OPT}}$ that swaps at most $\ell' = 4\ell + 4  + \ell + 1 \geq m + \ell + 1$ centers. We now bound the cost of such an optimum solution $\mathcal{W}_{\textrm{OPT}}$ to be at most $3$ times the cost of an optimum clustering of $P^{(\ell+1)}$. Let $\mathcal{V}(v)$ be the points in $P^{(\ell+1)}$ that are closest to center $v\in \mathcal{V}$ in the clustering $\mathcal{V}$. We say that a center $u \in \mathcal{U}$ is \emph{good} if 
\begin{itemize}
    \item it forms a well-separated pair with a center in $\mathcal{V}$; and
    \item $\mathcal{V}(v) \subseteq P^{(0)}$, i.e., the center $v$ is not closest to any of the new points in the considered epoch.
\end{itemize}
The number of centers of $\mathcal{U}$ that are bad, i.e.,  not good,  is at most the number of centers that do not form well-separated pairs ($m$ many) plus the number of centers that form a well-separated pair with a center $v\in \mathcal{V}$ such that $\mathcal{V}(v) \cap \left( P^{(\ell+1)} \setminus P^{(0)} \right) \neq \emptyset$ (at most $\ell+1$ many). Denote the good centers of $\mathcal{U}$ by $u_1, u_2, \ldots, u_g$ and index the centers of $\mathcal{V} = \{v_1, v_2, \ldots, v_k\}$ so that $u_i$ form a well-separated pair with $v_i$ for $i\in \{1, \ldots, g\}$. Then the clustering $\{u_1, \ldots, u_g, v_{g+1}, \ldots, v_k\}$ is obtained from $\mathcal{U}$ by doing $k-g \leq m + \ell+1 \leq \ell'$ swaps. It follows that the cost of $\mathcal{W}_{\textrm{OPT}}$ is bounded by the cost of this solution and so
\begin{align*}
    \opt{soda}{& \hphantom{\leq{}}} \cost\left(\mathcal{W}, P^{(\ell+1))}\right) \opt{soda}{\\} & \leq 13\cdot \cost\left(\mathcal{W}_{\textrm{OPT}}, P^{(\ell+1))}\right)  \\
    & \leq 13 \cdot\left(  \sum_{i=1}^g \cost(u_i, \mathcal{V}(v_i)) + \sum_{i=g+1}^k \cost(v_i, \mathcal{V}(v_i))\right) \\
    & \leq 13 \cdot\left(  \sum_{i=1}^g 3 \cdot \cost(v_i, \mathcal{V}(v_i)) + \sum_{i=g+1}^k \cost(v_i, \mathcal{V}(v_i))\right) \\
    & = 39 \cdot \cost\left(\mathcal{V}, P^{(\ell+1)}\right)\,,
\end{align*}
where the first inequality follows from \cref{thm:lpswap} and the last inequality is due to the following lemma which we prove in \cref{sec:wellsep_cost}. 
\begin{lemma}
    \label{lem:wellsep_cost}
    If $u\in \mathcal{U}, v\in \mathcal{V}$ form a well-separated pair and $\mathcal{V}(v) \subseteq P^{(0)}$, 
    \begin{align*}
        \cost(u, \mathcal{V}(v)) \leq 3 \cdot \cost(v, \mathcal{V}(v))\,.
    \end{align*}
\end{lemma}

Now \epochalgo obtains $\mathcal{U}^{(\ell+1)}$  by calling \makeHOAlg on centers in $\mathcal{W}$. 
We bound the cost of $\mathcal{U}^{(\ell+1)}$ by the following lemma.
\begin{lemma}\label{lemma:cost-any-robust}
For any clustering $\mathcal{W}$, let $\mathcal{U}$ be the result of $\algrobustify(\mathcal{W})$ procedure described in \cref{sec:epoch} for a set of points $P$. We have
$$
\cost(\mathcal{U}, P) \leq \frac{3}{2} \cost(\mathcal{W}, P).
$$
\end{lemma}
\begin{proof}
By \cref{lem:makerobustatmostonce}, \algrobustify makes at most one call to \makeHOAlg per center.  Let $w_1, \ldots w_k$ denote the centers of $\mathcal{W}$ and let $w'_1, \ldots, w'_k$ denote the centers of $\mathcal{U}$ where $w'_j$ was obtained by a call to \makeHOAlg on $w_j$ or $w'_j = w_j$ if no such call was made. Further index the centers in the order in which the calls to \makeHOAlg were made, putting those centers last for which  no call was made.

With this notation, \epochalgo calls $\makeHOAlg$ on center $w_j$ with input parameter $t_j$ selected to be the smallest integer so that 
\begin{align*}
    10^{t_j} \geq \dist(w_j, \{w'_1, \ldots, w'_{j-1}, w_{j+1}, \ldots, w_k\})/100\,.
\end{align*}
This implies that $10^{t_j} \leq \dist(w_j, w'_i)/10$ for any $i<j$ and  
$10^{t_j} \leq \dist(w_j, w_i)/10$ for any $i>j$. \cref{lem:nested_balls} says that $\dist(w_j, w'_j) \leq 10^{t_j}/2$. Hence, for two centers $w_i$ and $w_j$ with $i< j$,
\begin{align*}
    10 \cdot 10^{t_j} \opt{soda}{&} \leq  \dist(w'_i, w_j) \leq \dist(w_i, w_j) + \dist(w_i, w_i') \opt{soda}{\\ &} \leq \dist(w_i, w_j) + 10^{t_i}/2 \leq  \frac{11}{10} \dist(w_i, w_j)\,,
\end{align*}
where the last inequality follows as by the properties of the ordering, i.e., $t_i \leq t_j$. If we let $\mathcal{W}(w_j)$ be the points of $P$ assigned to center $w_j$ in the clustering $\mathcal{W} = \{w_1, w_2, \ldots,  w_k\}$, it follows that $\Ball(w_j, 10^{t_j}) \subseteq \mathcal{W}(w_j)$ since $10^{t_j} < \dist(w_j, w_i)/2$ for any $i\neq j$.  We can therefore apply \cref{lem:robust_cost} (with $P' = \mathcal{W}(w_j)$) to bound the increase of the cost by
\begin{align*}
    \cost(\mathcal{U}, P)
    \opt{soda}{&} = \sum_{j=1}^k \cost(w'_j, \mathcal{W}(w_j))
    \opt{soda}{\\ &} \leq \frac{3}{2} \sum_{j=1}^k \cost(w_j, \mathcal{W}(w_j))
    = \frac{3}{2} \cost(\mathcal{W}, P) \,.
\end{align*}
\end{proof}
By the above lemma, we conclude 
\begin{align*}
    \cost\left(U^{(\ell+1)}, P^{(\ell+1)}\right)
    \opt{soda}{&} \leq \frac{3}{2} \cdot \cost\left(\mathcal{W}, P^{(\ell+1))}\right)
    \opt{soda}{\\ &} \leq 100\cdot \cost\left(\mathcal{V}, P^{(\ell+1)}\right)\,,
\end{align*}
i.e., that $U^{(\ell+1)}$ is a $100$-approximate solution.  As aforementioned this implies that $U^{(\ell+1)}$ is  bounded-robust which in turn implies \cref{thm:maincost}. It remains to prove \cref{lem:well_separeted_remove} and \cref{lem:wellsep_cost} which we do in \cref{sec:number_wellseparated} and \cref{sec:wellsep_cost}, respectively.

\subsection{Proof of \cref{lem:well_separeted_remove}:  Relating Non-Well-Separated and Removed Centers}
\label{sec:number_wellseparated}

Let $k-m$ be the number of well-separated pairs in $\mathcal{U}$ and $\mathcal{V}$. We prove \cref{lem:well_separeted_remove}, i.e., that there is a clustering $\mathcal{U}' \subseteq \mathcal{U}$ obtained from $\mathcal{U}$ by removing at least $\lfloor m/4 \rfloor$ centers   and whose cost is bounded by
   \begin{align*}
       \cost\left(\mathcal{U}', P^{(0)}\right) \leq 6\gamma \left(\cost\left(\mathcal{U}, P^{(0)}\right) + \cost\left(\mathcal{V}, P^{(0)}\right)\right)\,.
   \end{align*}
   
   On a very high level, the reason why the above holds is that if centers are not too far away from each other then we can safely close some of them. The formal proof strategy is as follows. We define a feasible solution $(y, x)$ to the standard LP relaxation with the potential center locations $\mathcal{U}$ and point set $P$ such that
   \begin{itemize}
       \item  $\sum_{u\in \mathcal{U}} y_u \leq k - m/4$; 
       \item the cost is bounded by $\sum_{u\in \mathcal{U},p\in P} x_{up} \dist(u,p) \leq 2\gamma \left(\cost\left(\mathcal{U}, P^{(0)}\right) + \cost\left(\mathcal{V}, P^{(0)}\right)\right)$.
   \end{itemize}
   To this end, let us recall the standard LP relaxation for the weighted $k$-median problem. In the following LP, $w(p)$ represents the weight of the point $p$ for any $p \in P$,
   \begin{align}
    \min & \sum_{u\in \mathcal{U},p\in P}  x_{up} \cdot w(p) \cdot \dist(u,p) & \\
    \text{s.t.}  & \quad x_{up} \leq y_{u}  & \forall u\in \mathcal{U},p\in P \label{eq:open-connect0}\\
    &\sum_{u \in \mathcal{U}} x_{up} \geq 1  & \forall p \in P \label{eq:all-open0}\\
    &\sum_{u\in\mathcal{U}} y_u \leq k-\frac{m}{4} & \\
    &x_{up}, y_u \geq 0 & \forall u\in \mathcal{U},p\in P \label{eq:non-negative0}
\end{align}
   \cref{lem:well_separeted_remove} then follows because the integrality gap of the standard LP relaxation is known to be at most $3$~\cite{DBLP:conf/approx/CarnesS11}.
   We proceed to describe the fractional opening of the  centers (the setting of $y$), the fractional assignment of points to centers (the setting of $x$), and we then analyze the cost of the solution.
   
   \paragraph{Fractional Opening of Centers.}
    Let $\pi: \mathcal{V}  \rightarrow \mathcal{U}$ be the function that maps each center $\mathcal{V}$ to its closest center in $\mathcal{U}$ (breaking ties arbitrarily). Partition $\mathcal{U}$ into $\mathcal{U}_I$ and $\mathcal{U}_F$, where
        $\mathcal{U}_I$ contains those centers $u \in \mathcal{U}$ that satisfies at least one of the following conditions:
        \begin{itemize}
            \item $u$ forms a well-separated pair with one center in $\mathcal{V}$;
            \item $|\pi^{-1}(u)| \geq 2$.
        \end{itemize}
        So $\mathcal{U}_F$ contains those centers $u\in \mathcal{U}$ that does \emph{not} form a well-separated pair and $|\pi^{-1}(u)| \leq 1$.  We  define a solution to the standard LP in which every center  $u\in \mathcal{U}_I$ is integrally opened $y_u=1$ and each center $u\in \mathcal{U}_F$ is fractionally opened $y_u = 1/2$. 
        
        We now show that $\sum_{u\in \mathcal{U}} y_u \leq k - m/4$. Each center $u \in \mathcal{U}$ that forms a well-separated pair with a center $v\in \mathcal{V}$ has $|\pi^{-1}(u)| \geq 1$ since $u$ must be the closest center to $v$ in $\mathcal{U}$. In addition, as there are $k-m$ well-separated pairs in total, we get
        \begin{align*}
            k \geq \sum_{u\in \mathcal{U}_I} | \pi^{-1}(u)| \geq k- m   + 2\cdot\left(|\mathcal{U}_I| - (k-m)\right)\,,
        \end{align*}
        and so $|\mathcal{U}_I| \leq k - m/2$. We thus have
          $\sum_{u\in \mathcal{U}} y_u \leq k-m/2 + 1/2 \cdot m/2 = k - m/4$ as required.

   \paragraph{Fractional Assignment of Points.}     
        Each point $p\in P$ is assigned as follows. Let $v_p\in \mathcal{V}$ be the closest center to $p$ in $\mathcal{V}$ and let $u_p = \pi(v_p)$ be the closest center in $\mathcal{U}$ to $v_p$. We distinguish three cases:
        \begin{itemize}
            \item If $y_{u_p} = 1$ then assign point $p$ to $u_p$, i.e., set $x_{u_pp} = 1$. This assignment costs $w(p) \cdot \dist(u_p, p)$.
            \item If $y_{u_p} = 1/2$ and there is a center $u'_p\in \mathcal{U} \setminus \{u_p\}$ such that $\dist(u_p, u'_p) \leq \gamma\cdot  \dist(u_p, v_p)$, then assign $p$ equally to $u_p$ and $u_p'$, i.e., set $x_{u_pp} = x_{u_p' p} = 1/2$ (see \cref{fig:ca-lp2}). This assignment costs 
            \begin{align*}
                \opt{soda}{& \hphantom{\leq {}}} \frac{1}{2} w(p) \left(\dist(p, u_p) + \dist(p, u'_p)\right) \opt{arxiv}{&}
                \opt{soda}{\\ &} \leq \frac{1}{2} w(p) \left( \dist(p, u_p) + \dist(p, u_p) + \dist(u_p, u_p') \right) \\
                &  \leq w(p) \left( \dist(p, u_p) + \frac{\gamma}{2} \dist(u_p, v_p)\right)\,.
            \end{align*}
            \item If $y_{u_p} = 1/2$ and the previous case does not hold, then since $u_p, v_p$ does not form a well-separated pair, there is a center $v'_p \in \mathcal{V} \setminus \{v_p\}$ such that $\dist(v_p, v'_p) \leq \gamma \cdot \dist(u_p, v_p)$ (see \cref{fig:ca-lp3}). We assign $p$ equally to $u_p$ and $u'_p = \pi(v'_p)$, i.e., we set $x_{u_p p} = x_{u'_p p} = 1/2$. This assignment costs
            \begin{align*}
                \opt{soda}{& \hphantom{\leq {}}} \frac{1}{2} \left(\dist(p, u_p) + \dist(p, u'_p)\right) \opt{arxiv}{&}
                \opt{soda}{\\ &} \leq \frac{1}{2} w(p) \big( \dist(p, u_p) + \dist(p, u_p) \opt{soda}{\\ & \qquad} + \dist(u_p, v'_p)  + \dist(v'_p, u'_p)\big) \\
                &\leq \frac{1}{2} w(p)\big( \dist(p, u_p) + \dist(p, u_p) \opt{soda}{\\ & \qquad} + \dist(u_p, v'_p) +   \dist(v'_p, u_p)\big) \\
                &= w(p) \left( \dist(p, u_p) +  \dist(u_p, v'_p) \right) \\
                & \leq w(p) \left(\dist(p, u_p) + \dist(u_p, v_p) + \dist(v_p, v'_p)\right) \\
                &\leq w(p) \left(\dist(p, u_p) + (\gamma + 1) \dist(u_p, v_p)\right)\,.
            \end{align*}
        \end{itemize} 
        In the third case, we have $u_p \neq u_p'$ since $\pi(v_p) = u_p$ and $|\pi^{-1}(u_p)| \leq 1$ since $y_{u_p} = 1/2$ and therefore $u_p \in \mathcal{U}_F$. We have thus that each point is fractionally assigned one unit and that $x_{up} \leq y_u$ for all $u\in \mathcal{U}$ and $p\in P$. In other words, $x$ is a feasible fractional assignment of the points.

        \paragraph{Bounding the Cost.}  By the above calculations, the cost of assigning a point $p\in \mathcal{V}(v)$ is always at most
        $w(p) \left(\dist(p, u_p) + (\gamma+1) \dist(u_p, v_p)\right)$, with equality in the third case. 
           We thus have that the total cost is upper bounded by
           \begin{align*}
                 \opt{soda}{& \hphantom{\leq {}}} \sum_{p\in P} w(p)\left(\dist(p,u_p) + (\gamma+1) \dist(u_p, v_p)\right) \opt{soda}{\\}
                 &\leq 
                 \sum_{p\in P} w(p)\left(\dist(p,v_p) + (\gamma+2) \dist(u_p, v_p)\right) \\
                 &= \cost\left(\mathcal{V}, P^{(0)}\right) +  (\gamma+2)\sum_{p\in {P}} w(p) \dist(u_p, v_p)\,.        
             \end{align*}
             
             To bound the last term, for a point $p\in P$, let $u_p^*$ denote its closest center in $\mathcal{U}$.  Then by the definition of $u_p = \pi(v_p)$ and the triangle inequality,
             \begin{align*}
                 \dist(u_p, v_p) \leq \dist(u_p^*, v_p) \leq \dist (u^*_p, p) + \dist(p, v_p)\,.
             \end{align*}
             We thus have 
             \begin{align*}
                \opt{soda}{&} \sum_{p\in P} w(p) \dist(u_p, v_p) \opt{soda}{\\}
                \leq \opt{soda}{{}&{}} \sum_{p\in P} w(p)\left(\dist(u^*_p, p) + \dist(p, v_p)\right) \opt{soda}{\\}
                = \opt{soda}{{}&{}} \cost\left(\mathcal{U}, P^{(0)}\right) + \cost\left(\mathcal{V}, P^{(0)}\right)
            \end{align*}
            and the cost of the fractional solution is at most
             \begin{align*}
                 \opt{soda}{&} \cost\left(\mathcal{V}, P^{(0)} \right) + (\gamma+2)\Big(\cost\left(\mathcal{U}, P^{(0)}\right) \opt{soda}{\\ & \qquad} + \cost\left(\mathcal{V}, P^{(0)}\right) \Big) \opt{soda}{\\}
                 \leq \opt{soda}{{}&{}} 2\gamma \left(\cost\left(\mathcal{U}, P^{(0)}\right) + \cost\left(\mathcal{V}, P^{(0)}\right)\right)\,,
             \end{align*}
             as required.

\subsection{Proof of \cref{lem:wellsep_cost}: Cost of Well-Separated Pairs}
\label{sec:wellsep_cost}

In this section we prove \cref{lem:wellsep_cost}, i.e., the statement that
    if $u\in \mathcal{U}, v\in \mathcal{V}$  form a well-separated pair and $\mathcal{V}(v) \subseteq P^{(0)}$,  then
    \begin{align*}
        \cost(u, \mathcal{V}(v)) \leq 3 \cdot \cost(v, \mathcal{V}(v))\,.
    \end{align*}
    The statement is trivial if $\dist(u,v) = 0$ so we assume throughout that $\dist(u,v) \geq 1$.
   
   Since $\mathcal{U}$ is a bounded-robust solution, we have that $u$ is $t$-robust where $t$ is the smallest integer such that 
   \begin{align} \label{eq:dist-u-v-t}
      10^t \geq \opt{soda}{{}&{}} \dist(u, \mathcal{U}\setminus \{u\})/100 \opt{soda}{\\}
      \geq \opt{soda}{{}&{}} (\gamma/100) \cdot  \dist(u,v) =  20 \cdot  \dist(u,v)\,,
    \end{align}
    where the second inequality holds because $u$ and $v$ form a well-separated pair and the  equality is due to the selection of $\gamma=2000$.  Select $t^*$ to be the integer that satisfies  $20\cdot  \dist(u,v) \leq 10^{t^*}  < 10\cdot 20 \cdot  \dist(u,v)$. As $t^* \leq t$, there is a $t^*$-robust tuple $(p_0, p_1, \ldots, p_{t^*})$ with $t^*\geq 1$ and $p_0 = u$.   Let $B_i = \Ball(p_i, 10^i)$ for $i = 0,1, \ldots, t^*$. We shall use the following simple fact:
    
    \begin{claim}
        For every $i = 0,1, \ldots, t^*$, $B_i \subseteq \mathcal{V}(v)$.
    \end{claim}
    \begin{proof}[Proof of claim] Consider a point $q\in B_i$.  We have
    \begin{align*}
        \dist(q, v) & \leq \dist(q, p_i) + \dist(p_i, u) + \dist(u, v)  \\
        & = \dist(q, p_i) + \dist(p_i, p_0) + \dist(u,v)  \\ 
        & \leq 10^i + 10^i/2 + 10^{t^*}/20\\
        & \leq 2\cdot 10^{t^*}\,,
    \end{align*}
    where, for the penultimate inequality, we used $q\in B_i$, ~\cref{lem:robust_cost}, and the selection of $t^*$. That $q\in \mathcal{V}(v)$ and thus $B_i \subseteq \mathcal{V}(v)$ now follows because $u$ and $v$ form a well-separated pair and so by the selection of $t^*$ (specifically, that $10^{t^*} < 200 \dist(u,v)$)
    \begin{align*}
        \dist(v, \mathcal{V}\setminus \{v\}) \opt{soda}{&} \geq 2\cdot 10^3 \cdot  \dist(u,v) > 10^{t^* + 1} \opt{soda}{\\ &} \geq 2 \cdot \dist(q,v)\,. 
    \end{align*}
    \end{proof}
    We divide the remaining part of the proof of the lemma into two cases.
    Suppose first that $\avgcost(p_{t^*}, B_{t^*}) \geq 10^{t^*}/5$. We then have $p_{t^*-1} =  p_{t^*}$ and so 
    \begin{align}
        \dist(v, p_{t^*}) =  \dist(v, p_{t^*-1})  & \leq \dist(v, p_0) + \dist(p_0, p_{t^*-1})  \notag\\
        & = \dist(u, v) +  \dist(p_0, p_{t^*-1}) \notag\\
        & \leq 10^{t^*}/20 + 10^{t^*}/20 \notag\\
        & \leq 10^{t^*}/10\,, \label{eq:dist-v-ptstar}
    \end{align}
    where the penultimate inequality holds by the selection of $t^*$ and because $\dist(p_0, p_{t^*-1}) \leq 10^{t^*}/20$ by \cref{lem:robust_cost}.
    Hence, since $\avgcost(p_{t^*}, B_{t^*}) \geq 10^{t^*}/5 \geq 2 \cdot \dist(v, p_{t^*})$
    and $B_{t^*} \subseteq \mathcal{V}(v)$ (by the above claim), 
  \begin{align*}
      \opt{soda}{& \hphantom{\leq {}}} \cost(v, \mathcal{V}(v)) &
      \opt{soda}{\\ &} = \cost(v, B_{t^*}) + \cost(v, \mathcal{V}(v) \setminus B_{t^*}) \\
      & =  w(B_{t^*}) \cdot \avgcost(v, B_{t^*}) + \cost(v, \mathcal{V}(v) \setminus B_{t^*}) \\
      & \geq  w(B_{t^*}) \cdot \left(\avgcost(p_{t^*}, B_{t^*}) - \dist(v, p_{t^*})\right) \opt{soda}{\\ & \qquad} + \cost(v, \mathcal{V}(v) \setminus B_{t^*}) \\
      & \geq  w(B_{t^*}) \cdot \frac{\avgcost(p_{t^*}, B_{t^*})}{2} + \cost(v, \mathcal{V}(v) \setminus B_{t^*}) \\
      & \geq \frac{\cost(p_{t^*}, \mathcal{V}(v))}{2}\,, 
  \end{align*}
  where the last inequality holds because any point  $q\in \mathcal{V}(v) \setminus B_{t^*}$ has 
  \begin{align*}
    \dist(q, v) \geq \dist(p_{t^*}, q) - \dist(p_{t^*}, v) \opt{soda}{&} \geq 10^{t^*} - \dist(p_{t^*}, v) \opt{soda}{\\ &} \geq  \dist(p_{t^*}, v)
  \end{align*}
  where the first inequality follows by triangle inequality, second inequality follows since $q\in \mathcal{V}(v) \setminus B_{t^*}$, and last inequality holds due to~\cref{eq:dist-v-ptstar}.
  We can therefore conclude the proof of the lemma in this case by applying~\cref{lem:robust_cost} which says that $\cost(p_0 , \mathcal{V}(v)) \leq (3/2) \cdot \cost(p_t^*, \mathcal{V}(v))$.

It remains to consider when $\avgcost(p_{t^*}, B_{t^*}) < 10^{t^*}/5$. We then have $$p_{t^*-1} = \arg\min_{p\in B_{t^*}} \cost(p, B_{t^*}).$$  Therefore, since $B_{t^*} \subseteq \mathcal{V}(v)$ (by the above claim)
   \begin{align*}
       \opt{soda}{& \hphantom{\leq {}}} \cost(p_{t^*-1}, \mathcal{V}(v)) \opt{soda}{\\} & = \cost(p_{t^*-1}, B_{t^*}) + \cost(p_{t^*-1}, \mathcal{V}(v) \setminus B_{t^*}) \\
       & \leq \cost(v, B_{t^*}) +  \cost(p_{t^*-1}, \mathcal{V}(v) \setminus B_{t^*}) \,,
   \end{align*}
   where the last inequality follows since $v \in B_{t^*}$ from \cref{eq:dist-u-v-t}.
   To bound the second term, we use ~\cref{lem:robust_cost} to obtain
   \begin{align*}
       \dist(q, p_{t^*-1}) \opt{soda}{&} \geq \dist(q, p_{t^*}) - \dist(p_{t^*}, p_{t^*-1}) \opt{soda}{\\ &} \geq 10^{t^*} - 10^{t^*}/2  \qquad \mbox{for any $q\in \mathcal{V}(v) \setminus B_i$}\,,
   \end{align*}
   and 
   \begin{align*}
       \dist(v, p_{t^*-1}) &\leq \dist(v,u) + \dist(u, p_{t^*-1}) \\
       &= \dist(v, u) + \dist(p_0, p_{t^*-1}) \\
       & \leq 10^{t^*}/20 + 10^{t^*}/20 \leq 10^{t^*}/10\,.  
   \end{align*}
   Hence, 
   \begin{align*}
   \frac{\dist(p_{t^*-1}, q)}{\dist(v, q)} \opt{soda}{&} \leq \frac{\dist(p_{t^*-1}, q)}{\dist(p_{t^*-1}, q) -\dist(p_{t^*-1}, v)} \opt{soda}{\\ &} \leq  \frac{ 10^{t^*}/2}{10^{t^*}/2 - 10^{t^*}/10} \leq 2 
   \end{align*}
   and so 
   \begin{align*}
       \opt{soda}{&} \cost(p_{t^*-1}, \mathcal{V}(v)) \opt{soda}{\\} \leq \opt{soda}{{}&} \cost(v, B_{t^*-1}) +  \cost(p_{t^*-1}, \mathcal{V}(v) \setminus B_{t^*-1}) \opt{soda}{\\} \leq \opt{soda}{{}&} 2\cdot \cost(v,  \mathcal{V}(v))\,.
   \end{align*}
   We can thus also conclude this case by applying \cref{lem:robust_cost} which give us $\cost(u, \mathcal{V}(v)) \leq 3\cdot \cost(v, \mathcal{V}(v))$.

\section{Consistent $k$-Clustering Algorithm and Analysis}
\label{sec:everything_alg}
In this section we present our consistent algorithm. One of the ingredients used in this algorithm is Meyerson's sketch. In this work we use a modified version of the Meyerson's sketch presented in \cite{meyerson}. Intuitively, Multi-Meyerson's sketch is an algorithm that produces a weighted instance of size $k \cdot \text{polylog}(n, \Delta)$ that any constant approximate solution of this instance is also a constant approximate solution for the original instance. These properties are explained more precisely in the following theorem. Notice that in this theorem, the elements are inserted one by one and there is no need to have access to the entire stream at the beginning.

\begin{restatable}{theorem}{thmmeyerson}[Multi-Meyerson's procedure] \label{thm:meyerson}
Given a set of initial points $P'$ and a stream of insertions of points $\mathcal{\sigma'} = \langle\sigma'_1, \sigma'_2, \ldots \sigma'_{t'}\rangle$ such that 
\[
18\OPT(P') \geq \OPT(P' \cup \{\sigma'_1, \sigma'_2, \ldots \sigma'_{t'}\}),
\]
Multi-Meyerson's procedure finds a weighted initial solution $P$ and a stream of insertions of weighted points $\mathcal{\sigma} = \langle \sigma_1, \sigma_2, \ldots \sigma_{t}\rangle$. The stream is built on the fly, and satisfies the following properties with probability at least $1-\frac{1}{(n+\Delta)^{10}}$.
\begin{enumerate}
    \item The length of the produced stream $\mathcal{\sigma}$ is small, $$t \leq k \cdot \text{polylog}(n, \Delta).$$
    \item For any $1 \leq j \leq t'$, let $ \langle\sigma_1, \sigma_2, \ldots \sigma_i \rangle $ be the points on the stream $\sigma$ after the $j$-th insertion of stream $\sigma'$. Let $\mathcal{U}$ be a set of centers inducing an $\alpha$-approximate solution for the weighted instance defined by $P \cup \{\sigma_1, \sigma_2, \ldots \sigma_i \}$, then $\mathcal{U}$ is also a $O(\alpha+1)$-approximate solution for $P' \cup \{\sigma'_1, \sigma'_2, \ldots \sigma'_{j}\}$. I.e.,
    \begin{align*}
        \opt{soda}{&} cost(\mathcal{U}, P' \cup \{\sigma'_1, \sigma'_2, \ldots \sigma'_{j}\}) \opt{soda}{\\}
        \leq \opt{soda}{{}&{}} O(\alpha+1)\cdot\OPT(P' \cup \{\sigma'_1, \sigma'_2, \ldots \sigma'_{j}\}).
    \end{align*}
    \item The weights are positive integers and the total sum of the weights is at most $2n$.
\end{enumerate}
\end{restatable}
A full description of the Multi-Meyerson procedure along with a formal proof for this theorem is presented in \opt{arxiv}{\cref{sec:meyerson}}\opt{soda}{the full version of this paper}.

Our algorithm first runs the Multi-Meyerson procedure, therefore we need to detect when the cost of the optimum solution increases. To that end, independent from Meyeson's sketch we compute an $\alpha$-approximate solution (for $\alpha \leq 3$) for the entire instance after each insertion. Afterwards when it increases by a factor $6$, we get that the value of the optimum solution has increased by a factor at least $2$ and at most $18$. In this case restart the Multi-Meyerson procedure. This results in at most $\log (n\Delta)$ restarts of this algorithm and guarantees the condition in \cref{thm:meyerson} is satisfied. 

At the beginning of each run of the Multi-Meyerson procedure we compute a bounded-robust solution by simply finding a $3$-approximate solution for the the initial points produced by the Multi-Meyerson procedure. Then we make it robust, as explained in \cref{sec:epoch}. By \cref{lemma:cost-any-robust} we lose a factor $\frac{3}{2}$ in this step, therefore the resulting solution is a $\frac{9}{2}$-approximate solution. We pass it to \epochalgo as the initial solution. We keep running \epochalgo until the stream $\mathcal{\sigma}$ is finished, where each call to it uses the last solution created by the previous call. Notice that we know that the final solution created by \epochalgo is bounded-robust. Let us now analyze the approximation ratio and the total number of the changes in the solution.

\paragraph{Approximation Ratio.} Notice that the only place that we output a solution is in \epochalgo which we showed that is a constant approximate algorithm in \cref{main-epoc-cost}. Combining it with \cref{thm:meyerson}, we get that the solution produced is indeed a constant approximate solution throughout the algorithm.

\paragraph{Consistency of the Solution.}
At the beginning of each run of Multi-Meyerson, we compute a new solution which might result in $k$ changes compared to the previous solution. Therefore the total changes caused by it is at most $k \log (n\Delta)$. The rest of the changes happens during the \epochalgo, and by \cref{thm:epoc-consistency} we get that the total number of changes is at most a factor $O(\log^2 \Delta)$ more than the total length of the streams which is $k \cdot \text{ polylog}(n, \Delta)$ by \cref{thm:meyerson}, therefore we conclude that the total number of the changes is at most $k \cdot \text{polylog}(n, \Delta)$.

\paragraph{Success Probability.}
The algorithm is successful if no error happens in the following three steps.
\begin{enumerate}
    \item Swapping the centers: The probability that this step fails is also at most $\frac{1}{(n+\Delta)^{10}}$ from \cref{thm:lpswap} and is called at most $O(n \log(n\Delta))$ times.
    \item Multi-Meyerson procedure: The probability that this step fails is at most $\frac{1}{(n+\Delta)^{10}}$ from \cref{thm:meyerson} and is called at most $O(\log (n\Delta)^2)$ times.
\end{enumerate}

Putting together these two properties results in
\mainresult*

\paragraph{Removing the Assumption that $\Delta$ and $n$ are Known in Advance.} Throughout this paper, we assumed that we know the value of the $\Delta$ and $n$ in advance. This knowledge is not necessary in order to run our algorithm. Notice that we can keep track of both $n$ and $\Delta$ during the execution of the algorithm. When one of them increases by a constant factor, we restart our algorithm from the scratch. This results losing a constant factor in approximation guarantee, a polylog$(n, \Delta)$ factor in total number of changes in consecutive center sets, and the success probability becomes $1-O(\frac{1}{n^8})$.
\section{LP-based Algorithm for Swapping and Removing Centers} \label{sec:lp-algo}
In this section we design two LP-based algorithms for the following tasks:

\paragraph{Swapping Centers.}
Compute a set of $k$ centers that differs from an input set of centers in at most $O(\ell)$ centers and such that the cost of their induced clustering is a constant-factor approximation to the best possible clustering whose centers differ in at most $\ell$ from the input set of centers. 
\paragraph{Removing Centers.} Compute a minimal subset of the input centers that achieves a constant-factor approximation to the input instance.

Throughout this section we assume that the instance is not weighted and we copy the points according to their weights. %
Notice that from \cref{thm:meyerson} we know that the weights of the points are positive integers with total summation of at most $2n$.
\subsection{LP-based Algorithm for Swapping Centers} 
\label{sec:swapping-centers}

In this section we show%
\begin{restatable}{theorem}{thmswappingcenters} \label{thm:lpswap}
There exists an algorithm that given a set of centers $\mathcal{U}$, a set of points $P$ and a parameter $\ell$, finds a new set of centers $\mathcal{V}$ with probability at least $1-\frac{1}{(n+\Delta)^{10}}$, such that
\begin{enumerate}
    \item These two sets of centers differ in at most $4\ell$ centers, i.e.,
    $$ |\mathcal{U} \setminus \mathcal{V}| \leq 4 \ell.$$
    \item The cost of $\mathcal{V}$ is at most a factor $13$ larger than the best set of centers $\mathcal{W}$ that differs with $V$ in at most $\ell$ centers.
    $$\cost(\mathcal{V}, P) \leq 13 \cdot \min_{\mathcal{W}:  |\mathcal{U} \setminus \mathcal{W}| \leq \ell }  \cost(\mathcal{W}, p).$$

\end{enumerate}
\end{restatable}
To this end, we us the following LP-relaxation, where $y_0$ is equal to one if $i \in \mathcal{U}$ and zero otherwise. 
\begin{align}
    \min & \sum_{i,j \in P} x(i,j) \dist(i,j) & \label{eq:objective}\\
    \text{s.t.}  & \quad x(i,j) \leq y(i)  & \forall i,j \in P \label{eq:open-connect}\\
    &\sum_{i \in P} x(i,j) \geq 1  & \forall j \in P \label{eq:all-open}\\
    &\sum_{i\in P} y(i) \leq k \label{eq:k-open} \\
    &v(i) \geq y(i) - y_0(i) & \forall i \in P \label{eq:centerchange} \\
    &\sum_{i\in P} v(i) \leq \ell \label{eq:totalchange}\\ 
    &v(i), x(i,j), y(i) \geq 0 & \forall i,j \in P \label{eq:non-negative}
\end{align}

Intuitively, \cref{eq:objective} minimizes the cost of the solution.  \cref{eq:open-connect} ensures that a center is open before a point is assigned to it. \cref{eq:all-open} ensures that all the point are assigned to a center. \cref{eq:k-open} ensures that there are at most $k$ open centers. \cref{eq:centerchange} and \cref{eq:totalchange} ensure that the solution found is at most a value $\ell$ different from the initial solution $y_0$. Together with \cref{eq:non-negative} that ensure all the variables are positive, gives that   
\[
\sum_{i\in P} \max\{0,y(i) - y_0(i)\} \leq \ell.
\]

We refer to the optimum solution of the above LP as $y^*, x^*$ and $w^*$ and let \OPT denote its cost. One can observe that for any set of centers $\mathcal{W}$ such that $|\mathcal{U} \setminus \mathcal{W}| \leq \ell$, we have 
\begin{align} \label{eq:lpopt-cost}
    \cost(\mathcal{W}, P) \geq \OPT.
\end{align}
 Afterwards we use the rounding procedure described in \cite{DBLP:conf/icalp/CharikarL12} over the $y^*$ and $x^*$ fractional solution and obtain a random solution $y$ and $x$. Therefore, we get that
\begin{lemma}[Restated from \cite{DBLP:conf/icalp/CharikarL12}]\label{lem:charikar}
There is a rounding procedure that opens at most $k$ centers and the probability that $i$ is open is exactly $y^*(i)$. Moreover,  in expectation, the cost of the solution is at most a factor $3.25$ more than the fractional solution $x^*$.
\end{lemma}
Using Markov's inequality, we get that with probability at least $3/4$ 
\[
\sum_{i\in P} \max\{0,y(i) - y_0(i)\} \leq 4\ell,
\]
and similarly we get that with probability at least $3/4$ 
\[
\cost(y, P) \leq 3.25\cdot 4 \sum_{i,j \in P} x^*(i,j) \dist(i,j) \leq 13 \cdot \OPT.
\]
Combining with \cref{eq:lpopt-cost} we get that for any set of centers $\mathcal{W}$ such that $|\mathcal{U} \setminus \mathcal{W}| \leq \ell$,
$$\cost(y, P) \leq 13 \cost(\mathcal{W}, P).$$
Therefore, by applying the union bound we get that the  solution has both above properties with probability at least $1/2$. By repeating the rounding algorithm $O(\log (n+\Delta))$ times, we can boost the probability to $1-\frac{1}{(n+\Delta)^{10}}$.

\subsection{LP-based Algorithm for Polynomial Time Removing Centers}
\label{sec:removing-centers}
In this section we show that
\begin{restatable}{theorem}{thmremovingcenters} \label{thm:lpremove}
There exists an algorithm that given a set of centers $\mathcal{U}$, a set of points $P$ and a parameter $c > 1$, finds a set of centers $V$ such that
\begin{enumerate}
    \item $\mathcal{V} \subseteq \mathcal{U}$ such that
    $$\cost (\mathcal{V}, P) \leq 3c \cdot \cost(\mathcal{U}, P).$$
    \item For any set of centers $\mathcal{W} \subseteq \mathcal{U}$ such that $\cost (\mathcal{W}, P) \leq c \cdot \cost(\mathcal{U}, P)$, we have  $$|\mathcal{V}| \leq |\mathcal{W}|.$$
\end{enumerate}
\end{restatable}
We use the classic LP for the k-median problem with parameter $\ell$ that controls the size of the solution.
\begin{align}
    \min & \sum_{i \in \mathcal{U}, j \in P} x(i,j) \dist(i,j) & \label{eq:objective2}\\
    \text{s.t.}  & \quad x(i,j) \leq y(i)  & \forall i \in \mathcal{U}, \forall j \in P \label{eq:open-connect2}\\
    &\sum_{i \in \mathcal{U}} x(i,j) \geq 1  & \forall j \in P \label{eq:all-open2}\\
    &\sum_{i \in \mathcal{U}} y(i) \leq k-\ell \label{eq:k-open2} \\
    &x(i,j), y(i) \geq 0 & \forall i,j \in P \label{eq:non-negative2}
\end{align}

Intuitively, \cref{eq:objective2} minimizes the cost of the solution.  \cref{eq:open-connect2} ensures that a center is open before a point is assigned to it. \cref{eq:all-open} ensures that all the point are assigned to a center. \cref{eq:k-open} ensures that there are at most $k-\ell$ open centers. \cref{eq:non-negative} ensures that all the variables are positive.

Given a value $c$, we try all possible $k$ values of $\ell$ and return the largest one such that the cost of the above LP is at most a factor $c$ away from the cost of the optimum solution (denoted by $\OPT$). After the value $\ell$ is fixed, we can use any algorithm for the $k$-median problem with the number of centers equal to $k-\ell$, set of point $P$ and set of potential centers $\mathcal{U}$.  Therefore we get a $3$-approximate solution $\mathcal{V}$ such that, $|\mathcal{V}| \leq k-\ell$ and $\cost(\mathcal{V}, P) \leq 3 \OPT \leq 3c \cdot\cost(\mathcal{U}, P)$.

We know that there exist no fractional solution that has size less than $k-\ell$ and cost less than $c \cdot \cost(\mathcal{U}, P)$, therefore there exist no such a integral solution as well. This means that any solution $\mathcal{W} \subseteq \mathcal{U}$ such that $\cost (\mathcal{W}, P) \leq c \cdot \cost(\mathcal{U}, P)$, we get that $|\mathcal{W}| \geq k-\ell$. Therefore, $$|\mathcal{V}|  \leq k-\ell \leq |\mathcal{W}|\,.$$

\bibliographystyle{abbrv}
\bibliography{lit}
\appendix

\section{Multi-Meyerson Procedure}
\label{sec:meyerson}
We first present Meyerson's sketch which depends on an estimate of the optimum solution.  Our approach is logically the same as \cite{meyerson} but not identical. Then we prove that it has the desired guarantees in expectation,  and use Markov inequality to achieve that the properties are preserved with constant probability. The guarantees needed here are slightly stronger than the ones used in previous works, but the proof techniques are similar. Next, we boost the probability to $1-1/n^{10}$ by running copies of Meyerson's sketch. In this step we need to be extra careful with the weights of the points in the instance as it is important that the weights are only increasing. Specifically, we prove the following:
\thmmeyerson*

We describe the Multi-Meyerson Procedure and prove its correctness in the following subsections.

\subsection{Meyerson's Sketch}
Meyerson's sketching operates on a guess of an optimum solution denoted by $\GOPT$. We can use many of the known algorithms for the $k$-median problem to achieve a $\GOPT$ such that
$$\OPT(P') \leq \GOPT \leq 3\OPT(P').$$

Let us start by describing Meyerson's sketch and its analysis. For sake of simplicity, we add the point in the set $P' = \{p'_1, \ldots p'_{|P'|} \}$ to the beginning of stream $\sigma'$ in arbitrarily order resulting in 
$$
\sigma' = <p'_1, \ldots, p'_{|P'|}, \sigma'_1, \ldots, \sigma'_{t'}>. 
$$

\paragraph{Description of Meyerson's Sketch.} The algorithm maintains a solution $S$, i.e., a set of centers $S$ where initially $S=\emptyset$. Then the Meyerson's sketch as follows: 

\begin{description}
\item[\textnormal{\textit{Insertion of point $x$:}}]
We open $x$ with probability $h_x \leftarrow \min\{1,\dist(x,S) \frac{\klogn}{\GOPT}\}$ (where $\dist(x,\emptyset)$ is defined to be $\infty$). If $x$ is opened, we add it to $S$ and assign it to itself. Otherwise we assign $x$ to the closest open center in $S$ denoted by $f_x$. 
\end{description}

\paragraph{Analysis of Meyerson's Sketch.}
For completeness we start by analyzing the performance of Meyerson's Sketch. Note that the analysis is similar to the analysis presented in~\cite{meyerson}. We first bound the size of the dynamic set $S$ after all insertions, i.e., after insertion of $\sigma'_{t'}$. We let $\OPT_i$ denote the value of the optimum solution after $i$-th insertion for $1 \leq i \leq m$ and where $m$ is the length of the stream, i.e., $m = t' + |P'|$. Moreover, we let $P'_i$ and $S_i$ denoted the set of points and the solution maintained by the Meyerson's sketch after $i$-th insertion for $1 \leq i \leq m$. 

\begin{lemma}\label{lemma:sizes}
We have $$\E[|S_m|] \leq \beta k\,,\qquad 
\mbox{where }\beta =  \left(1+4\cdot\frac{\OPT_m}{\GOPT}\right) \klogn.
$$
\end{lemma}
\begin{proof}
Let $C^*$ indicate the clusters of an optimum solution after insertions, i.e., of an optimal clustering $P'_m$.  We partition the points according to $C^*$. Consider a center $c^* \in C^*$ and let $P(c^*)$ be the subset of points that is served by $c^*$ in the optimum solution. Moreover, recall that $\avgcost(c^*, P(c^*)) =\cost(c^*, P(c^*)) /|P(c^*)|.$
Now partition the points $P(c^*)$ as follows:
Let $P(c^*,0)  = \{x\in P(c^*) |  \dist(x,c^*) \leq  \avgcost(c^*, P(c^*)) \}$ be those points of $P(c^*)$ whose distance to $c^*$ is at most the average cost of the cluster,  and similarly, let $P(c^*,i)  = \{x\in P(c^*) | 2^{i-1} \avgcost(c^*, P(c^*)) < \dist(x,c^*) \leq  2^i \avgcost(c^*, P(c^*)) \}$. Notice that the total number of partitions that we have is $k(\log \Delta +1)$, since all the non zero pairwise distances are in $[1,\Delta]$. 
Now consider one of the cluster centers $c^*$ and a partition $P(c^*, i)$. The expected summation of $h_x$ value for $x \in P(c^*, i)$ before we open a center is one. Moreover, if a point $y_{(c^*,i)}$ is opened in this partition then for any point afterwards, $h_x \leq \dist(x,y_{(c^*,i)}) \frac{\klogn}{\GOPT}$. Let us bound $\dist(x,y_{(c^*,i)})$. By triangular  inequality we have: 
\begin{align} \label{eqxy}
    \dist(x,y_{(c^*,i)}) & \leq  \dist(x,c^*)+\dist(y_{(c^*,i)},c^*) \nonumber\\
                & \leq  \dist(x,c^*)+2\dist(x,c^*) + \avgcost(c^*, P(c^*))\nonumber\\
                & = 3 \dist(x,c^*)+ \avgcost(c^*, P(c^*)),
\end{align}
where the last inequality follows since by assumption $x$ and $y_{(c^*,i)}$ are always in the same partition, so if $x \in P(c^*, 0)$ then  $\dist(y_{(c^*,i)},c^*) \leq \avgcost(c^*, P(c^*))$ otherwise $\dist(y_{(c^*,i)},c^*) \leq 2\dist(x,c^*)$. Hence,
\begin{align*}
\E[|S|] &=  \E[\sum_{x \in P} h_x]\\
&\leq \sum_{x \in P} \E[h_x]\\
&\leq \sum_{(c^*, i)}\left(1+\sum_{x \in P(c^*,i)} \dist(x,y_{(c^*,i)}) \frac{\klogn}{\GOPT}\right)\\
&\leq \sum_{(c^*, i)}\left(1+\sum_{x \in P(c^*,i)} (3\dist(x,c^*)+\avgcost(c^*, P(C^*)))\frac{\klogn}{\GOPT}\right)\\
&\leq (\klogn + 4\OPT_m\frac{\klogn}{\GOPT})\\
&= \beta k\,.
\end{align*}
\end{proof}

We proceed to analyze the expected cost of the solution that is maintained by Meyerson's sketch. 
\begin{lemma}\label{lemma:cost}  
    For $1 \leq j \leq m$ we have
    $$\E[\sum_{x \in P'_j} \dist(x, f_x)]  \leq 4\OPT_j + \GOPT\,,$$
    where $f_x$ is set when $x$ is inserted as described above.
\end{lemma}
\begin{proof}
We also inherit the notations from the previous proof \cref{lemma:sizes}. Now consider one of the cluster centers $c^*$ and the partition $P(c^*,i)$.  Similar to the previous lemma the expected summation of the costs of points before we open a center is $\frac{\GOPT}{\klogn}$. Moreover, let $y_{(c^*,i)}$ be the first point opened in this partition then for any point inserted afterwards in this partition, $\dist(x,f_x) \leq \dist(x,y_{(c^*,i)})$. Hence,
\begin{align*}
\E[\sum_{x \in P'_j} \dist(x, f_x)]
&\leq \sum_{(c^*, i)}\sum_{x \in P(c^*,i)} \E[\dist(x,f_x)] \\
&\leq \sum_{(c^*, i)}\sum_{x \in P(c^*,i)} \E[\dist(x,y_{(c^*,i)})] \\
&\leq \sum_{(c^*, i)}\left(  \frac{\GOPT}{\klogn} +  \sum_{x \in P(c^*,i)} \left( 3\dist(x,c^*)+ \avgcost(c^*, P(c^*)) \right) \right) \\
& \leq \GOPT+(3\OPT_j+\OPT_j)\,,
\end{align*}
where the third inequality follows from \cref{eqxy}.
\end{proof}
\subsection{Extending Meyerson's Sketch to Multi-Meyerson}
In this section we show that we can achieve similar bounds as the previous section with high probability. To this end, we first show the properties that one can achieve by running copies of Meyerson's sketch. Afterwards we use this approach and present the Multi-Meyerson procedure. 

Recall that $\OPT_{|P'|} \leq \GOPT \leq 3 \OPT_{|P'|}$ and from the condition of \cref{thm:meyerson} we have $\OPT_m \leq 18 \OPT_{|P'|} $. In what follows, we assume that this condition is satisfied. Therefore we get the following two bounds on $\GOPT$.
\begin{observation} \label{obs:upper-gopt}
For $|P'| \leq j \leq m$, we have 
$$\GOPT \leq 6\OPT_j.$$
\end{observation}
\begin{proof}
This follows since adding new points can decrease the cost of the solution by at most a factor $2$, since it gives new options to open as centers. Therefore
$$\GOPT \leq 3 \OPT_{|P'|} \leq 3\cdot 2 \OPT_j \leq 6 \OPT_j.$$
\end{proof}
\begin{observation}\label{obs:lower-gopt}
$\GOPT \geq \frac{1}{18} \OPT_m.$
\end{observation}
\begin{proof}
Follows by combining $\GOPT \geq \OPT_{|P'|}$ and the assumption from \cref{thm:meyerson}.
\end{proof}

From \cref{obs:upper-gopt} and \cref{lemma:cost} we get 
\begin{lemma} \label{lemma:costprob}
For $|P'| \leq j \leq m$ with probability at least $3/4$
    $$\sum_{x \in P'_j} \dist(x, f_x) \leq 4(4\OPT_j + \GOPT) \leq 40 \OPT_j.$$.
\end{lemma}
\begin{proof}
The first inequality follows by applying markov inequality on \cref{lemma:cost}, and the second inequality follows from \cref{obs:upper-gopt}
\end{proof}

Similarly, from \cref{obs:lower-gopt} and \cref{lemma:sizes} by applying markov bound we get
\begin{lemma} \label{lemma:sizeprob}
With probability at least $3/4$, $$|S| \leq 4\beta k\,,\qquad 
\mbox{where }\beta =  (1+4\cdot18) \klogn.$$
\end{lemma}
In order to achieve high probability, we run $q = O(\log (n+\Delta))$ copies of Meyerson's sketch in parallel. Let $S^{(0)}_j, S^{(1)}_j, \ldots, S^{(q)}_j$ denote the set maintained by the copies after the $j$-th insertion of the stream $\sigma'$. Also let for any point $x$ the point that it's assigned to in the $i$-th copy be $f^{(i)}_x$.

Using \cref{lemma:sizeprob} and \cref{lemma:costprob}  
\begin{lemma} \label{lemma:sizecost}
With probability $1-\frac{1}{(n+\Delta)^{10}}$ for any $|P| \leq j \leq m$, there exist a $1 \leq i \leq q$ such that 
\begin{enumerate}
    \item $\sum_{x \in P'_j} \dist(x, f^{(i)}_x) \leq 40 \OPT_j.$
    \item $|S^{(i)}_j| \leq 292 \klogn.$
\end{enumerate}
\end{lemma}
\begin{proof}
The first property follows from \cref{lemma:costprob}. The second property follows from \cref{lemma:sizeprob} and the observation that $|S^{(i)}_j| \leq |S^{(i)}_m|$. The probability follows from the number of repetitions (simple product of complementary events).
\end{proof}
We are ready to explain the construction of the output set $P$ and the stream $\sigma$ and analyze its performance.
\paragraph{Multi-Meyerson Procedure.} We start with an empty stream $\sigma$ and additional vector of weights $w(x)$ where $w(x)$ is the weight of the point $x$. Notice that the weight of point $x$ when inserted to the stream is not necessary equal to $w(x)$, since $w(x)$ is dynamic and changes over the time but the weight of an inserted point cannot be changed. Moreover we might insert a point to the stream multiple times with different weights. For each point $x$ that is insert (on the stream $\sigma'$)\footnote{Recall that we added the point $P'$ to the beginning of the $\sigma'$.}, we append it to $\sigma$ with weight $1$ if it is added to $S^{(i)}$ for at least one $1 \leq i \leq q$ for which the current $|S^{(i)}| \leq 292\klogn$. In this case we let $w(x) = 1$ as well. Otherwise, we find the closest point $y \in \sigma$ to $x$ and increase the weight of $y$ by one. We assign $x$ to $y$ and let $f_x = y$. Now if $w(y) = 2^\ell$ for some value $\ell$, we insert $y$ to the stream with weight $2^\ell$. When the first $|P'|$ insertions are done, we let $P = \sigma$, clear $\sigma$ and continue the algorithm as before.  This step does not logically effect the algorithm. Basically it is dividing the created stream into initial solution and output stream.

\paragraph{Proof of \cref{thm:meyerson}:}
Let us start by bounding the length of the stream. We have $q$ copies of the Meyerson's sketch and each inserts at most $292\klogn$ centers with weight $1$ to the stream. The reason is that we skip adding the points added by the sketches with size more than $292\klogn$. Therefore in total we have $O(q\klogn)$ points with weight $1$. Moreover, each point that is inserted can be re-inserted at most $\log n+1$ times, since it is inserted at most once for each power of two and the highest power of two is at most $\log n$. Therefore we get that the length of the stream is upper bounded by
$$k\cdot \text{polylog}(n, \Delta).$$

We now analyze the quality of the solution. Let us focus on a value $r$, such that, $|P'| \leq r \leq m$ and fix the state of the algorithm and the variables after the $r$-th insertion of $\sigma'$ stream.  We first observe that, for any point $y$ on the stream, the number of points that are assigned to it is equal to $w(y)$, as we increase $w(y)$ anytime we assign a point to $y$. We let $v(y)$ be the summation of weights of the point $y$ currently on the stream. Select integer $\ell \geq 1$ such that $2^{\ell-1} \leq w(y) \leq 2^\ell-1$, by the description of the algorithm we get that
$$ v(y) = \sum_{1 \leq i < \ell} 2^i = 2^{\ell} -1.$$ 
Therefore, 
\begin{align} \label{eq:vw-weights}
    w(y) \leq v(y) \leq 2w(y).
\end{align}
This results in the third property in this theorem.

From \cref{lemma:sizecost} we get that with probability $1-\frac{1}{(n+\Delta)^{10}}$, there exist a copy of the Meyerson's sketch that has a good approximation ratio and a small size. Without loss of generality we assume that the copy is $S^{(0)}$ and for any $x$, the point that it's assigned to is $f^{(0)}_x$. Observe that size $|S^{(0)}| \leq 292 \klogn$, so all the points in $S^{(0)}$ are part of the stream. Therefore, $\dist(x, f_x) \leq \dist(x, f^{(0)}_x)$. The cost of each point $x$ is the cost to the closest center in $\mathcal{U}$ which is by triangular inequality at most
$$\dist(x, \mathcal{U}) \leq \dist(x, f_x) + \dist(f_x, \mathcal{U})\, .$$
Therefore the total cost is
\begin{align*}
\cost(\mathcal{U}, P'_r) &= \sum_{x\in P'_r} \dist(x, \mathcal{U}) \\
& \leq  \sum_{x\in P'_r} \dist(x, f_x) + \dist(f_x, \mathcal{U}) \\
& \leq  \sum_{x\in P'_r} \dist(x, f^{(0)}_x) + \dist(f_x, \mathcal{U}) \\
& \leq  \sum_{x\in P'_r} \dist(x, f^{(0)}_x) + \sum_{x\in P'_r} \dist(f_x, \mathcal{U}) \\
& \leq  40\OPT_r + \sum_{x\in P'_r} \dist(f_x, \mathcal{U})
\end{align*}
where the last inequality follows from \cref{lemma:sizecost}, now we use the weights to simplify the second term. To this end let $P_r = P \cup \sigma$, denote the union of the output set $P$ and current output stream $\sigma$. Recall that for any set $Q$, $\OPT(Q)$ is the optimum solution for the instance $Q$.
\begin{align*}
\cost(\mathcal{U}, P'_r) &= \sum_{x\in P'_r} \dist(x, \mathcal{U}) \\
& \leq  40\OPT_r + \sum_{x\in P'_r} \dist(f_x, \mathcal{U}) \\
& \leq  40\OPT_r + \sum_{y\in P_r} \dist(y, \mathcal{U}) \cdot w(y) \\
& \leq  40\OPT_r + \sum_{y\in P_r} \dist(y, \mathcal{U}) \cdot v(y) \\
& \leq  40\OPT_r + \alpha \cost(\OPT(P_r), P_r) \\
& \leq  40\OPT_r + \alpha \cost(\OPT(P'_r), P_r) \\
& \leq  40\OPT_r + \alpha \sum_{y\in P_r} \dist(y, \OPT(P'_r)) \cdot v(y) \\
& \leq  40\OPT_r + \alpha \sum_{y\in P_r} \dist(y, \OPT(P'_r)) \cdot 2w(y) \\
& \leq  40\OPT_r + \alpha \sum_{y\in P_r} \sum_{x \mid f_x = y} 2 (\dist(f_x,x) + \dist(x, \OPT(P'_r))) \\
& \leq  40\OPT_r + \alpha\sum_{x \in P'_r} 2 (\dist(f_x,x) + \dist(x, \OPT(P'_r))) \\
& \leq  (2\alpha+1)40\OPT_r + \alpha \sum_{x \in P'_r} 2\dist(x, \OPT(P'_r)) \\
& \leq  (2\alpha+1)40\OPT_r + 2\alpha \OPT_r \\
&\leq 82\alpha+40\OPT_r \, ,
\end{align*}
where the second inequality follows by rearranging the terms, the third inequality follows from \cref{eq:vw-weights}, the fourth inequality follows from \cref{thm:meyerson} since the clustering $\mathcal{U}$ is $\alpha$-approximate, the fifth inequality follows since the cost of any clustering for points in $P_r$ is more than the cost of an optimum clustering $\OPT(P_r)$ for it, the sixth inequality follows by definition of cost, seventh inequality follows by \cref{eq:vw-weights}, eighth inequality follows by triangle inequality and rearranging the terms.
This concludes the analysis of the quality of the solution and therefore the entire proof.

\end{document}